\documentclass[pra, twocolumn, superscriptaddress, groupaddress, floatfix]{revtex4-2}
\usepackage{slashed}
\usepackage{graphicx}
\usepackage{subfigure}
\usepackage[colorlinks=true,linkcolor=blue,citecolor=blue,urlcolor=blue]{hyperref}
\usepackage{amsmath}
\usepackage{amsfonts}
\usepackage[toc]{appendix}
\usepackage{txfonts}
\usepackage[british]{babel}
\usepackage[none]{hyphenat}
\usepackage{ulem}

\newcommand{\pr}[1]{\ensuremath{\left[#1\right]}} 
\newcommand{\pc}[1]{\ensuremath{\left(#1\right)}} 
\newcommand{\md}[1]{\ensuremath{\left\vert#1\right\vert}} 
\newcommand{\av}[1]{\ensuremath{\left\langle#1\right\rangle}} 
\DeclareMathOperator{\Real}{Re}
\DeclareMathOperator{\Imag}{Im}

\DeclareMathOperator{\hH}{\hat{H}}
\DeclareMathOperator{\ha}{\hat{a}}
\DeclareMathOperator{\hb}{\hat{b}}
\DeclareMathOperator{\hbnewbase}{\hat{S}}

\DeclareMathOperator{\hD}{\hat{D}}

\DeclareMathOperator{\hP}{\hat{P}}

\DeclareMathOperator{\hq}{\hat{q}}

\begin{document}
\title{Influence of direct dipole-dipole interactions on the optical response of 2D materials in strongly inhomogeneous infrared cavity fields}
\author{Sofia Ribeiro}
\affiliation{Material Physics Center, CSIC-UPV/EHU, \\ Paseo Manuel de Lardizabal 5, 20018 San Sebasti\'an-Donostia, Basque Country, Spain}
\author{Javier Aizpurua}
\author{Ruben Esteban}
\affiliation{Material Physics Center, CSIC-UPV/EHU, \\ Paseo Manuel de Lardizabal 5, 20018 San Sebasti\'an-Donostia, Basque Country, Spain}
\affiliation{Donostia International Physics Center (DIPC), \\ Paseo Manuel de Lardizabal 4, 20018 San Sebasti\'an-Donostia, Basque Country, Spain}
%
\begin{abstract}
A two-dimensional (2D) material, formed for example by a self-assembled molecular monolayer or by a single layer of a van der  Walls material, can couple efficiently with photonic nanocavities, potentially reaching the strong coupling regime. The coupling can be modelled using classical harmonic oscillator models or cavity quantum electrodynamics Hamiltonians that often neglect the direct dipole-dipole interactions within the monolayer. Here, we diagonalize the full Hamiltonian of the system, including these direct dipole-dipole interactions. The main effect on the optical properties of a typical 2D system is simply to renormalize the effective energy of the bright collective excitation of the monolayer that couples with the nanophotonic mode. On the other hand, we show that for situations of extreme field confinement, large transition dipole moments and low losses, fully including the direct dipole-dipole interactions is critical to correctly capture the optical response, with many collective states participating in it. To quantify this result, we propose a simple equation that indicates the condition for which the direct interactions strongly modify the optical response.
\end{abstract}
\maketitle
\section{Introduction}
Light and matter couple strongly when a large number of molecules, a van der  Waals material or a similar system, is placed within nanophotonic cavities \cite{PhysRevB.67.085311(2003), ACSPhot1.130(2015)}. In the strong coupling regime, the wavefunctions of the photonic modes and the material excitations mix to form new hybridized collective states known as polaritons \cite{NatComm6.5981(2015), JPhysChemLett6.1027(2015), ACSPhot10.1460(2015), ACSPhot1.130(2015), AngChem58.10635(2019)}. 
For example, the strong coupling between photonic modes and electronic molecular transitions results in the formation of excitonic polaritons \cite{Nature395.53(1998), PhysRevLett82.3316(1999), Science288.1620(2000)}. Moreover, there is growing interest in vibrational polaritons that are due to the coupling of vibrational modes of molecules and infrared (IR) microcavities 
\cite{NatComm.9.4012(2018), PhysRevLett.114.196403(2015),JPhysChemLett.7.354(2016), ACSNano2.2292(2021)}.
The emergence of these new IR polaritonic states can significantly impact the physical and chemical properties of the system \cite{ACS5.205(2018), PNAS114.3026(2017), PhysRevLett116.238301(2016), AngChem58.15324(2019), Science363.615(2019), PhysRevLett.114.196403(2015), PhysRevX.5.041022(2015)}, allowing for active manipulation of matter. 

The optical properties of  two-dimensional (2D) systems located in a cavity can be studied using cavity Quantum Electrodynamics (c-QED), or via classical harmonic oscillators models \cite{QEDbook, RepProgPhys78.013901(2014)}, which often neglect the direct dipole-dipole coupling between the different polarizing units, such as the molecules or the different regions of the material (unit cells). Within this framework, the system typically shows two optically-bright polariton modes under strong enough coupling strength, whether we have one or an ensemble of polarizable units forming a 2D material. The energy of these polaritonic modes is different from those of the uncoupled nanocavity and the vibrational or electronic excitation in the material, and their energy difference (Rabi splitting) increases with the number $N$ of polarizable units, as $\sim \sqrt{N}$ in simple situations  \cite{ACS49.2403(2016), JChemPhys155.050901(2021)}. Additionally, $N-1$ dark modes are also present that interact much more weakly with cavity photons, or not at all \cite{PNAS115.4845(2018)}.

On the other hand, the direct dipole-dipole interactions between the polarizing units of the 2D material can also influence the optical response. In a simple example, applying the Clausius-Mossotti equation to an ensemble of molecules (or oscillators) indicates that the resonances of the classical permittivity are shifted from the energy of the individual oscillators \cite{IntNanophotonics}. This effect is considered implicitly, for example, in a recent work studying the coupling between collective lattice vibrations (phonons) in hexagonal boron nitride (hBN) and microcavity modes \cite{NatComm12.6206(2021)}; the hBN layer was treated as an ensemble of dipoles, with resonant energy defined by the classical permittivity, which served to take into account the dipole-dipole interactions in an effective manner. In a similar context, it has been shown in Refs.~\cite{PhysRevA93.063835(2016), PhysRevLett116.233601(2016)} that a dense atomic cloud could be described as an homogeneous particle with an effective permittivity. In these  works, the authors show that there is a correspondence between the microscopic polaritonic modes of the atomic cloud -- obtained by considering dipole-dipole interactions between the atoms --, and the modes of a homogeneous particle.

In this paper, we use a microscopic c-QED description of the dynamics of excited states to gain further insights into the effect of the direct dipole-dipole interaction on the optical response. We focus on the coupling of a nanophotonic cavity mode with vibrations of a 2D material (see Fig.~\ref{Fig:Setup}), which could consist of a self-assembled molecular monolayer or a single layer of a van der Waals material. Van der  Waals materials manifest very clear phonon modes with large reststrahlen bands that enable new optical properties \cite{Science6309.1992(2016), LightSciAppl7.17172(2018), Nature562.557(2018), Science6386.291(2018)}, while molecular monolayers have applications in the design of different devices such as chemical sensors, biosensors, and organic field-effect transistors (see Ref.~\cite{RSC8.3938(2020)} and references within). For simplicity, we often refer below directly to molecular assemblies, even if some values used for the parameters can be more representative of van der  Walls materials.
The cavity fields can present extreme spatial confinement, down to sub-nanometer regions, as occurs in plasmonic cavities formed by atomic-size protrusions (picocavities) in, e.g., Scanning Tunneling Microscopy, Nanoparticle-on-Mirror constructs or similar configurations \cite{Nature498.82(2013), NanoLett5.3410(2015), Science6313.726(2016), Science6399.251(2018)}. We emphasize that our methodology is very general, and the conclusions can be applied directly to other related situations, including the coupling with excitonic molecular transitions.
 
In the following, we first describe in Sec.~\ref{sec:theory} a general theory of the coupling between vibrations in the 2D material and a nanocavity mode. The model treats the vibrations as point-like dipoles, with each dipole corresponding to a molecule (or to a microscopic region of the 2D material, such as a unit cell \cite{NatComm12.6206(2021), arxiv2204.09641}). The vibrations occur in the direction perpendicular to the 2D monolayer. In Sec.~\ref{sec:collective}, we diagonalize the vibronic Hamiltonian in the absence of a nanophotonic cavity to obtain the new eigenmodes of the system, corresponding to the collective vibrational modes. We then include the cavity mode in Sec.~\ref{sec:coupling} and write the full interaction Hamiltonian as a function of these collective vibrational modes, which we solve to find the new vibron-polariton modes. Based on the properties of these modes, we obtain and analyze the optical response of the coupled system to reveal the effect of the direct dipole-dipole interaction. Finally, in Sec.~\ref{sec:conclusions}, we give concluding remarks.

\section{Theoretical Model \label{sec:theory}}
\begin{figure}[t]
    \centering
    \includegraphics[width=0.95\columnwidth]{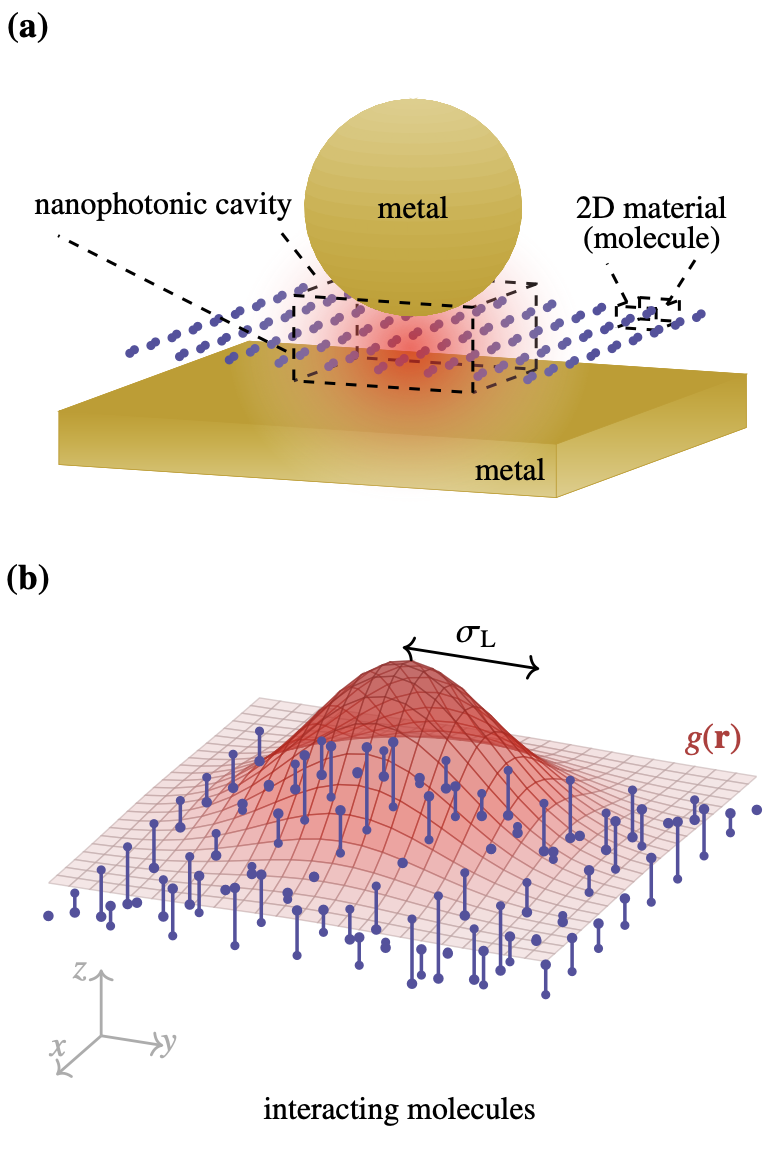}
    \caption{
    (Color online) System under study. 
    \textbf{(a)} Schematic of an ensemble of $N$ molecules placed in a 2D square lattice configuration inside a photonic (plasmonic or phononic) nanocavity. The cavity fields are represented by the red-shaded area.
    \textbf{(b)} Schematic of the excitation of the molecules by the nanocavity and of the collective mode emerging from the direct dipole-dipole coupling between the molecules. The inhomogeneous cavity IR field varies as a function of position $\mathbf{r}$ following a Gaussian distribution $g(\mathbf{r})$ of width (standard deviation) $\sigma_\text{L}$, as represented by the red-colored surface. If the molecules interact with each other, a collective, or cooperative vibronic behavior of the 2D molecular vibrations can emerge. Each resulting collective mode can then be seen as a density wave along the 2D plane forming a standing-wave pattern. The vertical lines illustrate the induced molecular dipoles corresponding to one such density wave. The schematics are not to scale.
    \label{Fig:Setup}}
\end{figure}

We consider a patch of $N$ molecules (polarizing units in the 2D-material) that form a monolayer inside an IR nanophotonic cavity (e.g., a plasmonic or phononic nanoresonator), as sketched in Fig.~\ref{Fig:Setup}~(a). For our numerical simulations, we arrange $N=51 \times 51 = 2601$ molecules in a lattice of square unit cell with lattice constant $a$ (that we fix at $a = 0.5$~nm), placed in the $xy$-plane. Each molecule is modeled as a dipole associated to a molecular vibration that is optically active at IR frequencies. The same model is also suited to treat other excitations, such as electronic transitions. The dipoles are oriented perpendicularly to the $xy$-plane, along the $z$-direction. We consider perfectly regular arrays with dipoles oriented along a fixed direction, but we expect that small randomness in the position or orientation of the molecules would affect the results weakly.

If the molecules are sufficiently far apart from each other, the molecular vibrations (or the excitonic transitions) are typically assumed to interact much more efficiently with the localized cavity field than directly with each other. This localized field can be confined to a very small region \cite{NanoLett5.3410(2015)}, as represented schematically by the red-colored surface in Fig.~\ref{Fig:Setup}~(b), strongly increasing the coupling between the cavity and the molecules \cite{Nature498.82(2013), Science6313.726(2016), PhysRevLett.118.127401(2017)}. However, when the molecules are closely packed, the direct dipole-dipole interaction between molecules can be important. For example, if the dipole-dipole Coulomb interaction between the molecules cannot be neglected, new collective modes emerge even in the absence of the nanophotonic cavity. These can be understood as collective density charge waves where the dipoles oscillate perpendicularly to the molecular plane with a characteristic pattern (i.e., a standing wave like pattern in the molecular plane with a characteristic in-plane wavevector resulting from the interference between the density waves due to reflection at the edges of the molecular patch) -- these are localized 2D phonon polaritons.
One such collective mode is exemplified by the vertical lines in  Fig.~\ref{Fig:Setup}~(b), with the length of each line illustrating the strength of the dipole associated with the vibration of the corresponding molecule collectively forming the stationary pattern. 

To study the optical response of this system, we consider the Hamiltonian describing the interaction of one IR nanocavity mode with the ensemble of molecular vibrations, which also interact with each other,
\begin{align}
    \hH &=
    \hH_\text{pht} + \hH_\text{mol} + \hH_\text{vib-pht} + \hH_\text{vib-vib}
    \nonumber\\
    &=
    \hbar\, \omega_\text{cav} \ha^\dagger \ha + \hbar \sum_{j = 1}^N \omega_j \hb_j^\dagger \hb_j + \hbar \sum_{j = 1}^N  g_{j } \pc{\ha^\dagger + \ha} \pc{\hb^\dagger_{j}+ \hb_{j}} \nonumber \\
    &\quad + \hbar \sum_{j} \sum_{l>j} \Omega_{jl} \pc{\hb^\dagger_{j } + \hb_{j }}
\pc{\hb^\dagger_{l} + \hb_{l}},
    \label{eq:HvibphtDip}
\end{align}
where $\ha$ and $\ha^\dagger$ are the bosonic creation and annihilation operators of the nanocavity excitations (e.g., plasmons or phonon polaritons), with frequency $\omega_\text{cav}$, respectively. The molecular vibrations at frequency $\omega_j$ are quantized using the vibron (i.e., a quantum of intramolecular vibration) creation and annihilation operators $\hb_j$ and $\hb_j^\dagger$, where the index $j$ distinguishes between the molecules. Thus, the terms $\hH_\text{pht}$ and $\hH_\text{mol}$ in eq.~\eqref{eq:HvibphtDip} correspond to the energy of the cavity mode and that of the vibrations, respectively. The third term ($\hH_\text{vib-pht}$) describes the interaction between the vibration of each molecule and the IR cavity field mode, with coupling strength $g_j$. 
Here, we specifically consider a nanocavity mode with a Gaussian field distribution (Fig.~\ref{Fig:Setup}~(b)). Thus, for identical molecules the coupling strength is proportional to the cavity field,
\begin{align}
    g_j \pc{\mathbf{r}_j} = g_{0} \, \exp \left[-\frac{(x_j-x_0)^2+(y_j-y_0)^2}{2 \sigma_\text{L} ^2}\right],
\end{align}
with $|\mathbf{r}_j|=\sqrt{x_j^2+y_j^2}$, $g_{0}$ the value at the center $\mathbf{r}_0 = (x_0,\, y_0)$ of this Gaussian function, and $\sigma_\text{L}$ the standard deviation.

Finally, $\hH_\text{vib-vib}$ describes the intermolecular dipole-dipole interaction, where the coupling strength, $\hbar \, \Omega_{jl}$, between each pair of molecules $j$ and $l$ is given by the (static) Coulomb coupling as 
\begin{align}
    \hbar\, \Omega_{jl} = \frac{1}{4 \pi \varepsilon_0 r_{jl}^3}  \pr{\mathbf{d}_j \cdot \mathbf{d}_l - 3 \pc{\mathbf{d}_j \cdot \mathbf{e}_{jl}}\pc{\mathbf{d}_l \cdot \mathbf{e}_{jl}}} .
    \label{eq:p0}
\end{align}
Here, $\mathbf{d}_j$ is the (real) transition dipole moment vector of molecule $j$, $\mathbf{r}_{jl}$ the distance vector between molecule $j$ and $l$ (with $\mathbf{e}_{jl}$ its unit vector), $r_{jl} = \md{\mathbf{r}_{jl}}$, and $\varepsilon_0$ the vacuum permittivity.

We place the molecules in the $xy$-plane, as described previously, and consider that all molecules have the same transition dipole moment $\mathbf{d}_j \equiv \mathbf{d}_\text{mol}$ (aligned along the $z$-axis) and same bare energy $\omega_j \equiv \omega_\text{mol}$.
In the following numerical calculations, we parameterize the interaction by considering the coupling between nearest neighbors in a 1D chain,  $$\hbar\, \Omega_{jl} (r_{jl} = a) \equiv \hbar\, \Omega_0.$$ This model can be extended to different lattice configurations (including disordered ensembles), different dipole orientations, and to samples where the molecules differ from each other.

Furthermore, note that we do not include a diamagnetic term  $ \hH_\text{diam} =  \mathcal{D} \, (\ha^\dagger + \ha)^2$, that is often considered in USC, where $\mathcal{D} = \sum_j \md{g_j}^2 / \omega_j$ \cite{NatRevPhys1.19(2019)}. The effect of this term on the diagonalization of the Hamiltonian can be reproduced by shifting the cavity frequency,  $\omega'_\text{cav} \to \omega_\text{cav} + 2\mathcal{D}$ \cite{PhysRevA91.063840(2015)}. Thus, our conclusions should not be affected by the inclusion of this diamagnetic term.

\section{Collective vibrational states \label{sec:collective}}

In order to solve the full Hamiltonian of the system (eq.~\eqref{eq:HvibphtDip}) and to better understand the emergence of collective vibrational modes, we first neglect the nanocavity mode and diagonalize the vibrational contribution to the Hamiltonian $\hH_\text{coll}=\hH_\text{mol} +\hH_\text{vib-vib}$. Following the Bogoliubov procedure \cite{Bogoliubov}, there exists collective bosonic operators $\hP_{n}$ that are linear combinations of the  vibrational operators $\hb_j$ of the individual molecules,
\begin{align}
\hP_{n } = \sum_{j=1}^N \pc{\alpha_{n j} \hb_{j } + \beta_{nj} \hb^\dagger_{j}},
\label{eq:Pn}
\end{align}
and diagonalize $\hH_\text{coll}$ according to 
\begin{align}
\hH_\text{coll}=\hH_\text{mol} +\hH_\text{vib-vib} &= \sum_n \hbar\, W_{n} \hP^\dagger_{n} \hP_{n}.
\label{eq:Hcoll}
\end{align}
$\hP_{n }$ and $W_{n }$ are the new operators and eigenfrequencies of the collective modes of the system, and $ \pr{\hP_{n }, \hH_\text{coll}} = \hbar W_{n } \hP_{n }$. The subscript $n$ refers to the index of the collective modes and $j$ to the index of the molecules. 
To obtain the values of $W_{n }$ and the $\alpha_{n j}$ and $\beta_{nj}$ coefficients, we write
\begin{align}
\pr{\hP_{n},\hH_\text{mol} +\hH_\text{vib-vib}} &= \hbar\, W_{n} \hP_{n} =\hbar\, W_{n} \sum_{j=1}^N \pc{ \alpha_{nj} \hb_{j} + \beta_{nj} \hb^\dagger_{j}}.
\label{eq:6}
\end{align}
From eq.~\eqref{eq:Pn}, the left-hand side of this expression is also
\begin{align}
    \pr{\hP_{n},\hH_\text{mol} +\hH_\text{vib-vib}} &=
\sum_j \left\lbrace \alpha_{nj}  \pr{\hb_{j },\hH_\text{vib} +\hH_\text{vib-vib} }
\right. \nonumber\\
&\quad \left. + \beta_{nj}  \pr{\hb_{j }^\dagger,\hH_\text{vib} +\hH_\text{vib-vib} } \right\rbrace .
\label{eq:7}
\end{align}
Inserting the expression for $\hH_\text{mol} +\hH_\text{vib-vib}$ given by eq.~\eqref{eq:HvibphtDip} into the right-hand side of eq.~\eqref{eq:7},
and comparing the resulting equations with eq.~\eqref{eq:6}, we obtain a system of linear equations resulting in the eigenvalue problem $\mathbf{M} \mathbf{V}_{n } = W_{n } \mathbf{V}_{n }$, where $\mathbf{M}$ is the Hopfield matrix \cite{PhysRev112.1555(1958)}, 
\begin{align}
    \mathbf{M} &=
    \left( \begin{array}{ccccccc}
\omega_{1 } & 0 & \Omega_{12 } & - \Omega_{12 } 
& & \Omega_{1N } & - \Omega_{1N }
\\ 
0 & -\omega_{1 } & \Omega_{12 } & - \Omega_{12 }& \cdots & \Omega_{1N } & - \Omega_{1N }\\ 
\Omega_{12 } & -\Omega_{12 } & \omega_{2 } & 0 & & \Omega_{2N } & - \Omega_{2N }\\ 
\Omega_{12 } & -\Omega_{12 } & 0 & - \omega_{2 } & & \Omega_{2N } & - \Omega_{2N } \\ 
 & & \vdots & & \ddots & & \\
 \Omega_{1N } & -\Omega_{1N } & \Omega_{2N } & -\Omega_{2N } & & \omega_{N } &0\\
  \Omega_{1N } & -\Omega_{1N } & \Omega_{2N } & -\Omega_{2N } & & 0 & -\omega_{N }
 \end{array} \right)_{2N \times 2N} \!\!\!\!\!\!\!\!\!\! .
\end{align}
This matrix admits $2N$ eigenvalues, and if $W_n$ is an eigenvalue, so is $-W_n$. The $N$ distinct frequencies $W_n$ correspond to the normal modes, i.e., the collective states. The eigenvectors $\mathbf{V}_{n }$ are determined by the values of $\alpha_{nj}$ and $\beta_{nj}$  
\begin{align}
    \mathbf{V}_n^T = \left( \begin{array}{ccccc}
\alpha_{n1} & \beta_{n1} & \cdots & \alpha_{nN} & \beta_{nN} \end{array} \right)_{1 \times 2N },
\end{align}
where $\sum_{j=1}^N \pc{ \md{\alpha_{nj}}^2 - \md{\beta_{nj}}^2 } =1$ ensures the bosonicity of the operators.
The matrix $\mathbf{M}$ is real and block symmetric. Not all matrices $\mathbf{M}$ show exclusively real eigenvalues. However, for the parameters we consider, we find real eigenvalues and eigenvectors. Since the vector elements of $\mathbf{V}_n$ are real (i.e., $\alpha_{n j}^* = \alpha_{n j}$ and $\beta_{n j}^*=\beta_{n j}$), we can write from eq.~\eqref{eq:Pn}
\begin{align*}
    \hP^\dagger_{n } + \hP_{n } 
    &= \sum_{j=1}^N  \pc{\alpha_{n j} + \beta_{nj}} \pc{ \hb^\dagger_{j } + \hb_{j}},
\end{align*}
which can be written as a matrix-vector product. Inverting this matrix, allows us to write $\hb^\dagger_{j } + \hb_{j}$ as a function of $\hP^\dagger_{n } + \hP_{n }$.  
Defining the $N \times N$ inverse matrix $X_{j n} = \pc{\alpha_{n j} + \beta_{n j}}^{-1}$\cite{PRB86.125314(2012)}, we obtain
\begin{align}
\pc{ \hb^\dagger_{j }+ \hb_{j} }= \sum_n X_{j n} \pc{ \hP^\dagger_{n } + \hP_{n }}.
\label{eq:bvsP}
\end{align}

\subsection*{Analysis of the new collective vibrational modes}

\begin{figure}
    \centering
    \includegraphics{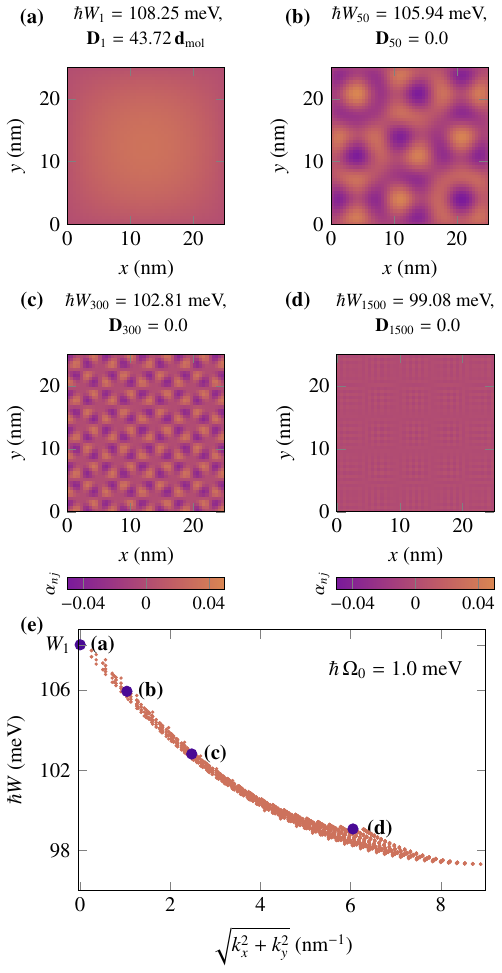}
    \caption{(Color online) Eigenmodes and eigenenergies of a molecular layer in the absence of nanocavity, for $N = 51 \times 51 = 2601$  molecules arranged in a 2D square lattice. 
    \textbf{(a)-(d)} Collective vibrational modes arising from dipole-dipole interactions. 
    The color plot shows the distribution of the values of the coefficient $\alpha_{nj}$ of the system eigenvectors in the $xy$-plane for the collective modes \textbf{(a)} $n=1$, \textbf{(b)} $n= 50$, \textbf{(c)} $n=300$ and \textbf{(d)} $n=1500$ (marked by the purple dots in \textbf{(e)}). Each $\alpha_{nj}$ is associated with a molecule, and thus with a spatial position. For each mode, we also indicate the corresponding eigenfrequency 
    $W_n$ and total dipole moment $\mathbf{D}_{n}$. 
    \textbf{(e)} Dispersion relation, corresponding to the energy of the modes of the system, plotted as a function of the wavevector $\md{\mathbf{k}} =\sqrt{k_x^2+k_y^2}$ associated with each eigenvector.
    In \textbf{(a-e)}, we use lattice constant $a=0.5$~nm,  dipole-dipole interaction $\hbar\, \Omega_{0} = 1~\text{meV}$ and vibrational energy $\hbar\, \omega_\text{mol} = 100~\text{meV}$.}
    \label{Fig:2DWjN}
\end{figure}
We show the nature of some illustrative collective modes in Fig.~\ref{Fig:2DWjN}~(a-d) for the $N=51 \times 51 = 2601$ molecules in a square lattice. 
In this paper, we have not considered any change of the interaction at edges, only the number of neighbors differs. We set the nearest-neighbor interaction strength to $\hbar \, \Omega_0  = 1~\text{meV}$  and indicate on top of each panel the corresponding eigenenergy $W_n$ of the mode $n$. The value of $\Omega_0$ is chosen to be comparable to the value for van der  Waals or other polar materials (see App.~\ref{App:Mol2Dmat}). Each panel represents the spatial pattern of the eigenvalue associated with the collective mode, which is plotted as a color plot of the amplitude of the $\alpha_{nj}$ coefficients (with each $\alpha_{nj}$ corresponding to a molecule $j$) as a function of the location of the molecules in the $xy$-plane. We observe that the collective mode with the largest energy (Fig.~\ref{Fig:2DWjN}~(a)) corresponds to all molecules oscillating in phase (same sign), while  other modes show approximately periodic changes of sign of the individual vibrational amplitudes along the $x$ and $y$-directions. Thus, the collective modes can be thought of as the standing wave pattern of a density wave of molecular vibrations (oscillating in the $z$-direction) with a characteristic in-plane wave vector along the $xy$-plane of the 2D molecular monolayer. The position of these collective modes in the energy dispersion relation of the molecular layer is marked as purple dots in Fig.~\ref{Fig:2DWjN}~(e).

To understand the coupling of the collective mode $n$ with light, it is useful to obtain the dipole moment of the collective mode $\mathbf{D}_n$, as only modes with a significant value of $\mathbf{D}_n$ couple efficiently, in the absence of nanocavity, with an incoming laser or other focused illumination (characterized by almost constant fields in the molecular ensemble). The total dipole can be written in terms of the operators of each molecule as $\hD = \sum_{j=1}^N \mathbf{d}_j \, (\hb^\dagger_j + \hb_j)$. Transforming to the collective picture, we can write
\begin{align}
    \hD = \sum_{j=1}^N X_{j n} \, \mathbf{d}_j \, (\hP^\dagger_n + \hP_n).
    \label{eq:Pntot}
\end{align}
We find that the collective mode with higher energy shows the highest non-zero total dipole moment, $\mathbf{D}_1=\sum_{j=1}^N \mathbf{d}_j X_{j 1} = 43.72\, \mathbf{d}_\text{mol}$. This efficient coupling in this mode can be expected because the induced molecular dipoles oscillate with the same phase, as mentioned previously. Lower energy eigenvalues have generally much lower values of $\mathbf{D}_n$ due to the sign changes of $\alpha_{nj} \, \pc{\beta_{nj}}$, and thus are more difficult to excite optically. For example, the low energy modes in Fig.~\ref{Fig:2DWjN}~(b-d) are characterized by  $\mathbf{D}_n=0$. 
To further characterize the collective modes, and based on the clear periodicity of the vibrational pattern in Fig.~\ref{Fig:2DWjN}~(a-d), we calculate for each mode $n$ the 2D Fourier transform of the  spatial $\alpha_{nj}$ maps in the $xy$-plane, $\md{\text{F.T.} \pr{\alpha_{nj}}}^2$.  We then define the characteristic wavevector of each mode, $k_{x,y}$, where $k_x$ and $k_y$ are the wavevector components in the $x$ and $y$-axis, respectively, as the value at which the corresponding maximum of $\md{\text{F.T.} \pr{\alpha_{nj}}}^2$ is found. In Fig.~\ref{Fig:2DWjN}~(e), we present the resulting dispersion relation, showing the energy of the modes as a function of the parallel wavevector $\sqrt{k^2_x+k^2_y}$. This dispersion follows closely that of an infinite 2D layer of identical molecules, as shown in App.~\ref{App1}. We find that the energies cover a frequency range between $97.32$~meV and $108.25$~meV. 
The results show a certain `spreading' of the data points, i.e. modes can have the same $\sqrt{\pc{k_x^\text{max}}^2 + \pc{k_y^\text{max}}^2}$, but a different energy because the square lattice is not isotropic, so  that the direction in the [$k_{x}$, $k_{y}$] plane influences the result, particularly for large $\sqrt{\pc{k_x^\text{max}}^2 + \pc{k_y^\text{max}}^2}$. Specifically, the obtained values are contained within an upper set of points corresponding to $k_y \,  (k_x) = 0  $ that reaches $k_{x}^\text{max} \, (k_{y}^\text{max}) = \pi / a \sim 6.28~\text{nm}^{-1}$, and a lower one for $ k_x^\text{max} = k_y^\text{max}$ that reaches $\sqrt{\pc{k_x^\text{max}}^2 + \pc{k_y^\text{max}}^2} = \sqrt{2} (\pi / a) \sim 8.89~\text{nm}^{-1}$ (see further discussion in App.~\ref{App1}).
As for our particular example, we have chosen to orientate the dipole moments in the direction perpendicular to the $xy$-plane where the molecules are situated, the modes of smaller wavevectors are characterized by larger energies.

\section{Collective vibrational dynamics of a molecular monolayer coupled to a cavity \label{sec:coupling}}

We consider next the effect of the coupling of the molecules with the nanophotonic mode. With this purpose, we return to the total Hamiltonian in eq.~\eqref{eq:HvibphtDip}, and use eq.~\eqref{eq:bvsP} to rewrite
\begin{align}
\hH_\text{vib-pht} &= \hbar\, \sum_{j } g_{j } \pc{\ha^\dagger + \ha} \pc{\hb^\dagger_{j}+ \hb_{j}} ,
\end{align}
in terms of the collective operators, which gives
\begin{align}
\hH_\text{pht-coll}
 &=\hbar \sum_n \sum_j  g_{j } X_{jn} \pc{\ha^\dagger + \ha} \pc{ \hP^\dagger_{n }+ \hP_{n} } \nonumber \\
 &=\hbar \sum_n  \mathcal{G}_n \pc{\ha^\dagger + \ha} \pc{ \hP^\dagger_{n }+ \hP_{n} },
\end{align}
where we defined $\mathcal{G}_n = \sum_j g_{j } X_{jn}$. Thus, in the new collective base
\begin{align}
    \hH &= \hbar\, \omega_\text{cav} \ha^\dagger \ha + \hbar \sum_n W_n \hP_n^\dagger \hP_n \nonumber\\
    &\quad + \hbar \pc{\ha^\dagger + \ha} \sum_n \mathcal{G}_n  \pc{ \hP^\dagger_{n }+ \hP_{n}}.
\end{align}

A photonic nanocavity mode characterized by homogeneous fields couples preferentially with the largest-energy mode due to its particularly large dipole strength (and the large resulting  $\mathcal{G}_n$). Thus, it is informative to diagonalize the Hamiltonian in the simple case that just one collective state of frequency $W$ interacts with one cavity mode. We look for the two new polariton operators $\hq_m= \zeta_{m1} \ha + \eta_{m1} \ha^\dagger + \zeta_{m2} \hP + \eta_{m2} \hP^\dagger$, where the Hopfield coefficients $\zeta_{lm}$, $\eta_{lm}$ ($l$, $m=1,2$) satisfy the normalization condition $\md{\zeta_{l1}}^2 - \md{\eta_{l1}}^2 + \md{\zeta_{l2}}^2-\md{\eta_{l2}}^2 = 1$. Following a Bogoliubov diagonalization procedure \cite{Bogoliubov}, the characteristic polynomial of the eigenproblem ($\det \pr{\mathbf{I} \mathcal{W} - \mathbf{M}}=0$) can be written as 
\begin{align}
\left(\omega_\text{cav} ^2-\mathcal{W}^2\right) \left(W_1^2-\mathcal{W} ^2\right)-4 \omega_\text{cav}  W_1 \, \mathcal{G}_1^2  =0.
\label{eq:characteristicpolynomial}
\end{align}
The solutions of this equation are the eigenenergies of the two vibron-polariton modes \cite{QEDbook},
\begin{align}
\mathcal{W}_\pm^2 &= \frac{1}{2} \Bigg( 
\omega_\text{cav} ^2+W_1^2 \nonumber \\ 
&\quad \pm \left. \sqrt{ \pc{\omega_\text{cav}^2-W_1^2}^2 
 + 16\, \mathcal{G}_1^2 \omega_\text{cav}  W_1} \right).
\label{eq:eigenvaluesonemolecule}
\end{align}
which indicates that the new modes are separated by an energy difference that depends on the coupling strength $\mathcal{G}_1$ and on the cavity detuning. Eq.~\eqref{eq:eigenvaluesonemolecule} is formally equivalent to the equation obtained for coupling between a single molecular vibration and a cavity mode \cite{Nature445.896(2007), RepProgPhys78.013901(2014)} (note that eq.~\eqref{eq:eigenvaluesonemolecule} slightly differs from the expression that is found when the rotating wave approximation is used). The main aspects to consider when coupling with a collective mode are: (i) $W_1$ in eq.~\eqref{eq:eigenvaluesonemolecule} corresponds to the frequency of the collective vibrational mode (without cavity), and not to the resonant frequency of the individual molecules, (ii) the coupling strength $\mathcal{G}_1$ includes a contribution from all molecules that couple with the photonic nanocavity \cite{PhysRevLett102.077401(2009),PhysRevB82.075429(2010)}, and (iii) together with the two bright modes at energies $\mathcal{W}_\pm$, there are $N-1$ dark modes that do not couple at all with the cavity in this approximation \cite{Optica10.1247(2018), PhysRevLett114.157401(2015), PNAS115.4845(2018)}. 

We consider next the general case where the nanophotonic cavity field can be strongly inhomogeneous, and we need to consider all the collective modes simultaneously. The  $N+1$ eigenfrequencies $\mathcal{W}_{m}$ and eigenvectors $(\zeta_{m1} \; \eta_{m1} \; \zeta_{m2} \; \eta_{m2} ...)^T$  of the system can be found by diagonalizing the matrix,
\begin{align}
\mathbf{M}=
 \left( \begin{array}{ccccccc}
\omega_\text{cav} & 0 & \mathcal{G}_{1} & - \mathcal{G}_{1} 
& & \mathcal{G}_{M} & - \mathcal{G}_{M}
\\ 
0 & -\omega_\text{cav} & \mathcal{G}_{1} & - \mathcal{G}_{1}& \cdots & \mathcal{G}_{M} & - \mathcal{G}_{M}\\ 
\mathcal{G}_{1} & -\mathcal{G}_{1} & W_{1 } & 0 & &0 & 0\\ 
\mathcal{G}_{1} & -\mathcal{G}_{1} & 0 & -W_{1} & & 0 & 0 \\ 
\vdots & \vdots & \vdots &\vdots & \ddots &\vdots &\vdots \\
 \mathcal{G}_{M} & -\mathcal{G}_{M} & 0 & 0 & & W_{M} &0\\
  \mathcal{G}_{M} & -\mathcal{G}_{M} & 0 & 0 &\cdots & 0 & -W_{M}
 \end{array} \right),
\end{align}
which results in the transformed Hamiltonian $\hH = \sum^{M}_{m=1} \hbar \, \mathcal{W}_{m} \hq_{m}^\dagger \hq_{m}$, where $M=N+1$.

To analyze the optical response of the system, we consider the typical situation where the optical dipole of the nanophotonic cavity is much larger than that of the molecules (or, in an alternative picture, that the field induced by the nanophotonic cavity is much larger than the incident field). Thus, the coupling of light with the system is mostly dominated by the photonic fraction of each mode $m$, which is given by $\pc{\md{\zeta_{m1}}^2 -\md{\eta_{m1}}^2}$. 
We then define the following spectral function,  
\begin{align}
    S(\omega) = \sum_m \pc{\md{\zeta_{m1}}^2 -\md{\eta_{m1}}^2} \frac{\gamma / 2}{(\omega-\mathcal{W}_m)^2 + (\gamma / 2)^2},
    \label{eq:SwNoInt}
\end{align}
to characterize the optical response of the system, where we include \textit{ad-hoc} the effect of losses through the damping rate $\gamma$, which were neglected in the original  Hamiltonian, converting the delta-like modes to Lorentzians. We set the losses to $\hbar\, \gamma = 1$~meV except when otherwise stated. For simplicity, eq.\eqref{eq:SwNoInt} does not consider interference effects between different modes that could induce Fano resonances under certain experimental conditions \cite{NanoLett8.2106(2008), NatMaterials8.758(2009), NatComm.9.4012(2018)}.
In App.~\ref{App:Losses}, we discuss how, for the systems considered here, introducing losses in this way gives almost identical results as including complex-valued (lossy) frequencies in the Hamiltonian.

\subsection*{Optical response of the vibrational modes}

\begin{figure}[t]
    \centering
    \includegraphics{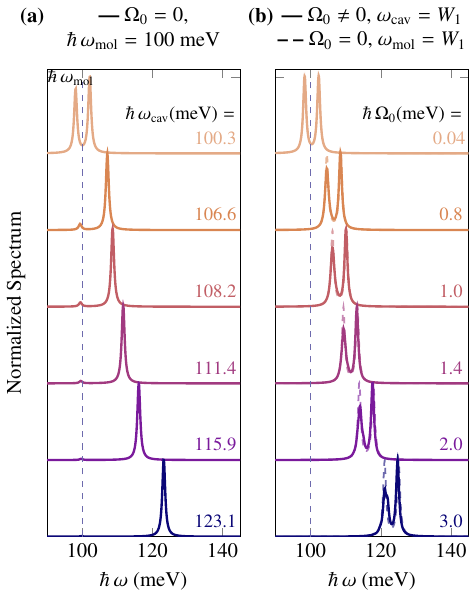}
    \caption{
    (Color online) Effect of direct molecule-molecule coupling on the optical spectra $S (\omega)$ of a spatially homogeneous IR cavity mode coupled to a 2D square lattice with $N=51 \times 51 = 2601$ molecules. The solid lines correspond to the normalized spectrum \textbf{(a)} ignoring $\pc{\Omega_0 = 0}$ and \textbf{(b)} including the molecule-molecule interaction $\pc{\Omega_0 \neq 0}$, with the resonant frequency of the molecule set to $\hbar \, \omega_\text{mol}= 100$~meV (highlighted by the light blue vertical dashed lines). 
    In \textbf{(a)} different spectra (shifted vertically for visibility) correspond to different frequencies of the cavity modes $\omega_\text{cav}$.
    In \textbf{(b)}, the same $\omega_\text{cav}$ are considered, and the dipole-dipole coupling strength  $\Omega_0$ is also changed between spectra. The $\Omega_0$ values are chosen so that, for each spectrum,  the maximum frequency of the collective eigenmodes (i.e. $W_1$, corresponding to the collective mode with associated wavevector $\mathbf{k} =0$ in Fig.~\ref{Fig:2DWjN}~(\textbf{a, e})) is equal to the cavity frequency. This condition corresponds to resonant coupling. The dashed lines in \textbf{(b)} correspond to the spectra obtained without the molecule-molecule interaction but setting $\omega_\text{mol} = W_1$, i.e. the frequency of the $\mathbf{k} =0$ collective mode that would be obtained if the molecule-molecule interaction was included). The results are obtained for a  homogeneous field distribution $\hbar\, g_j (\mathbf{r}_j) \approx 0.04$~meV, lattice constant $a=$ 0.5~nm, total molecules-cavity coupling strength  $\hbar\, g_\text{tot} = \hbar\, \sqrt{\sum_j \md{g_j (\mathbf{r})}^2}=2$~meV and $\hbar\, \gamma = 1$~meV.
    \label{Fig:FitCav}}
\end{figure}

For the following analysis, we study the effect of changing the different parameters: the strength of the dipole-dipole interaction $\Omega_0$, the confinement $\sigma_\text{L}$, the molecule-cavity strength $g_\text{tot}$, and the losses $\gamma$, separately. Other possibilities, such as modifying the lattice parameter $a$, can be thought as a combination of these separate changes (see more details in App.~\ref{App:Extra}).

We consider as a first step a nanophotonic cavity characterized by a homogenous field distribution ($\sigma_\text{L} \to \infty$), which couples with the 2D molecular square patch containing $N=51 \times 51 = 2601$ molecules that was analyzed in Sec.~\ref{sec:collective}. The optical spectrum of the system is shown in Fig.~\ref{Fig:FitCav}~(a) as a function of the frequency of the cavity $\omega_\text{cav}$ for the case of no direct molecule-molecule coupling $\Omega_0=0$ and molecule-cavity coupling strength $\hbar \, g_j (\mathbf{r}_{j}) \approx 0.04$~meV (through the manuscript, we fix $\hbar \, g_\text{tot} = \hbar \, \sqrt{\sum_j \md{g_j (\mathbf{r})}^2}$, except when stated otherwise).
For large detuning $\omega_\text{cav}-\omega_\text{mol}$, we observe a single peak at the frequency of the nanophotonic mode. As $\omega_\text{cav}\rightarrow \omega_\text{mol}$, the second mode becomes visible, and the two peaks show the typical avoided anti-crossing.

This avoided anti-crossing is a characteristic property of the strong coupling regime \cite{PhysRevLett58.353(1987), Nanoscale10.3589(2018)}, and can be  described by setting the dipole-dipole interaction to zero, $\Omega_{jl}=0$, and $\omega_j = \omega_\text{mol}$ in eq.~\eqref{eq:HvibphtDip} and defining $N$ new operators, $\hbnewbase_{j}=\sum_{l = 1}^N c_{jl} \hb_{l}$, where $c_{jl}$ are coefficients forming an orthonormal base and $\pc{c_{11}, \, c_{12}, \cdots, c_{1N}}= \pc{g_{1}, \, g_2, \cdots, g_N}/ \sqrt{ \sum_{j = 1}^N \md{g_{j }}^2}$ \cite{Optica10.1247(2018)}. $\hbnewbase_1$ is then the only collective mode coupled to the cavity, with a total effective coupling $$g_\text{tot} = \sqrt{ \sum_{j = 1}^N \md{g_{j }}^2},$$ thus allowing us to write
\begin{align}
    \hH_\text{coll} &=  \hbar\, \omega_\text{cav} \ha^\dagger \ha + \hbar\,  \sum_{j = 1}^N \omega_\text{mol}  \hbnewbase_j^\dagger \hbnewbase_j +\hbar\, g_\text{tot}   \pc{\ha^\dagger + \ha} \pc{\hbnewbase^\dagger_{1}+ \hbnewbase_{1}}.
    \label{eq:HamS1}
\end{align}
In Fig.~\ref{Fig:FitCav}, results for $\hbar \, g_\text{tot}= 2$~meV are shown. This Hamiltonian has $N+1$ excited eigenstates: two polariton modes, with eigenfrequencies $\approx \omega_\text{mol} \pm g_\text{tot}$  (resulting from the linear combination of the cavity mode and the collective bright state of the molecules), and $N-1$ dark modes that do not couple with the nanophotonic cavity (these dark modes are combinations of molecular excitations orthogonal to the collective bright state of the molecular excitation) \cite{PhysRevLett114.157401(2015), PNAS115.4845(2018)}. The exact expression (for no losses) of the eigenfrequencies of the polaritonic modes is given by eq.~\eqref{eq:eigenvaluesonemolecule} after changing $W \rightarrow \omega_\text{mol}$ and the effective coupling strength to $\mathcal{G}_1 \rightarrow g_\text{tot}$. 

The spectral response after switching on the molecule-molecule coupling, $\Omega_0 \neq 0$, is shown by the solid lines in Fig.~\ref{Fig:FitCav}~(b) for the same $\omega_\text{cav}$ values (we use a different value of $\Omega_0>0$ for each value of the cavity resonance, as indicated by the labels, for reasons explained below; since $\Omega_0$ is defined in units of energy, changing its value could be seen as changing $a$ while keeping the dipole moment unchanged, or vice-versa). In this case, $\omega_\text{cav}$, $\Omega_0>0$, we observe two closely situated peaks for all $\omega_\text{cav}$, which is drastically different from the results for the equivalent situation with $\Omega_0=0$ (and same values of $\omega_\text{cav}$). This difference can be connected with the dispersion of the collective modes for non-zero molecule-molecule coupling (Fig.~\ref{Fig:2DWjN}). The coupling with a cavity mode characterized by homogenous fields is dominated by the collective vibrational excitations of small wavevector $\md{\mathbf{k}}=\sqrt{k^2_x+k^2_y}$, whose eigenenergies are significantly larger than the molecular one. The energy of the excitations that couple with the cavity are thus very different depending on whether the direct molecule-molecule coupling is included or not. 

To take into account the shift from the molecular vibrational energy to the energy of the brightest collective mode, we also perform simulations where no molecule-molecule interactions are considered but setting the molecular frequency  to $\omega_\text{mol} = W_1$. $W_1$ corresponds to the maximum frequency in the dispersion, associated with the collective mode characterized by $\mathbf{k} = 0$ ($W_1$ is obtained for the values of $\Omega_0$ indicated by the labels, but the molecule-molecule interaction term in the Hamiltonian is not included in the calculation of the spectra). The obtained results are shown by the dashed lines in Fig.~\ref{Fig:FitCav}~(b), and are generally very similar to the results in Fig.~\ref{Fig:FitCav}~(b) with dipole-dipole interaction (and unshifted vibrational frequency $\hbar\, \omega_\text{mol} = 100$~meV), plotted with solid lines. Further, we note that the values of $\Omega_0$ for each cavity frequency $\omega_\text{cav}$ were chosen in Fig.~\ref{Fig:FitCav}~(b) so that $W_1=\omega_\text{cav}$ (resonant system), which explains why we always obtain two peaks of similar amplitude. 

The results in Fig.~\ref{Fig:FitCav}~(b) thus indicate that the molecule-molecule interaction can strongly modify the optical response but that, for a spatially homogenous cavity fields, this effect can be mostly corrected by considering a shifted vibrational frequency $\omega_\text{mol} \to W_1$. 
Shifts from the bare molecular frequencies in optical spectra due to dipole-dipole interactions can also occur in J-aggregate ensembles \cite{AngChIntEd15.3376(2011)} and absorbed molecules \cite{SpectActa7.1295(1965)}.

From a practical perspective, this renormalization of the energy can be accomplished by associating the vibrational frequency with the resonance of the classical permittivity of the infinite material (monolayer in the case in this work), as, to first approximation, the  classical permittivity is a response function that already contains the interactions between different regions of the material (via a Clausius-Mossotti-like relation for instance). This approach was followed, for example, in Ref.~\cite{NatComm12.6206(2021)}.

\begin{figure}[t]
    \centering
    \includegraphics{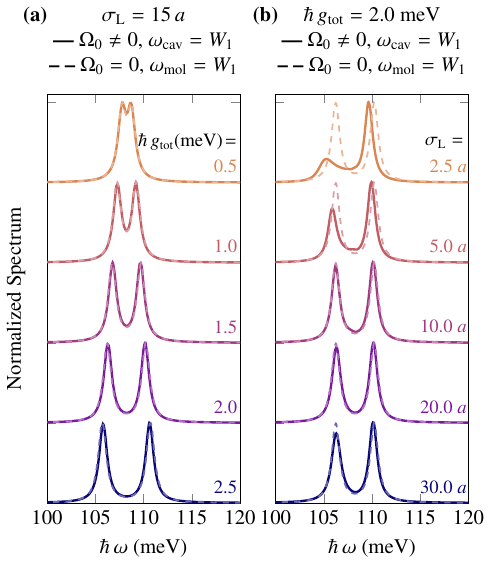}
    \caption{(Color online) Influence of the direct dipole-dipole interactions on the optical spectra $S (\omega)$ for different values of the coupling strength and field localization. The normalized optical spectra of an IR cavity coupled to a 2D square lattice with $N=51 \times 51 = 2601$ molecules is plotted \textbf{(a)} changing the total molecules-cavity coupling strength  $g^2_\text{tot} = \sum_j \md{g_j (\mathbf{r})}^2$  (values indicated in the figure) while keeping fixed the width of the field distribution to $\sigma_\text{L} = 15\, a$ and  \textbf{(b)} changing the field localization of the photonic mode $\sigma_\text{L}$ (values indicated in the figure) and keeping $\hbar \, g_\text{tot} =2$~meV fixed. Solid lines correspond to the results obtained  including direct molecule-molecule interactions with $\hbar\, \Omega_{0} = 1~\text{meV}$ and vibrational frequency $\hbar\, \omega_\text{mol}= 100$~meV. Dashed lines represent the values without molecule-molecule interactions and using $\hbar \, \omega_\text{mol} =\hbar \, W_1\approx 108.2$~meV. We also set $\hbar \, \omega_\text{cav} =\hbar \, W_1\approx 108.2$~meV for all spectra. The spectra are shifted vertically for visibility and are obtained for lattice constant $a=$ 0.5~nm. We set $\hbar \, \gamma = 1$~meV.
\label{Fig:Field}}
\end{figure}

We emphasize further the energy renormalization in  Fig.~\ref{Fig:Field}~(a) where we plot the results over a smaller frequency range for different effective coupling strength  $g_\text{tot} = \sqrt{\sum^N_{j=1} \md{g_j (\mathbf{r})}^2}$. In this case, we include a weak variation of the spatial field distribution of the photonic mode ($\sigma_\text{L} = 15 \, a$) and the system is in resonance $\hbar\, \omega_\text{cav} = \hbar\, W_1 \approx 108.2$~meV. Consistent with the results in Fig.~\ref{Fig:FitCav}, the results obtained including the molecule-molecule interactions $\hbar\, \Omega_0=1$~meV and using the vibrational frequency $\omega_\text{mol}$ (solid lines) are very similar to those obtained for $\Omega_0=0$ and $\hbar\, \omega_\text{mol} = \hbar\, W_1 \approx 108.2$~meV. All spectra show two almost symmetric peaks, with energy separation (Rabi splitting) that increases with growing $g_\text{tot}$. 

In contrast, the direct molecule-molecule interactions can strongly modify the optical spectra for a tightly confined cavity field. This effect is illustrated in  Fig.~\ref{Fig:Field}~(b), where we fix $\hbar\, g_\text{tot}= 2$~meV and vary the field confinement $\sigma_\text{L}/a$. The cavity resonance is again chosen to be resonant with $W_1$. The constant value of $g_\text{tot}$ assumed in this case implies a strengthening of the coupling of the cavity with each individual molecule, $g_j$, for increased field confinement (smaller $\sigma_\text{L}/a$), as the number of molecules  interacting with the cavity field is effectively reduced. We observe that, for $\sigma_\text{L}/a\lessapprox 5$, there is a striking difference between the results obtained including molecule-molecule interaction ($\hbar \, \Omega_0=1$~meV and $\hbar\, \omega_\text{mol} = 100$~meV, solid line) and those without these interactions ($\Omega_0=0$ and $\hbar\, \omega_\text{mol} = \hbar\, W_1 \approx 108.2$~meV, dashed line). The former shows a gradual smearing out and disappearance of the low-energy peak that is not present for the latter, a behavior that can be understood from the following simple picture. A weakly confined (i.e., almost uniform) nanocavity field couples preferentially with the single eigenmode at $\sqrt{k_x^2+k^2_y}=0$ that is characterized by a significantly larger dipole moment $\sum_{j} X_{jn} \,\mathbf{d}_\text{mol}$ than that of the other eigenvalues.  Thus, the optical response is dominated by the coupling between the photonic mode and one collective vibrational mode, resulting in the standard emergence of two almost symmetric peaks of a typical strongly coupled resonant system (see eq.~\eqref{eq:eigenvaluesonemolecule} and \eqref{eq:HamS1}). On the other hand, a strongly confined field can couple efficiently (large $\mathcal{G}_n$) with modes of higher order (larger $\sqrt{k_x^2+k^2_y}$), resulting in a more complex spectra with contributions from many collective vibrational modes along the dispersion curve.  As a subtle point, we note that the difference between the two spectra (with and without molecule-molecule interactions) is smallest for $\sigma_\text{L} / a\approx 10-20$ and it increases for less confined fields (larger $\sigma_\text{L}/a$), which we attribute to the illumination of the molecules near the edges of the molecular ensemble. However, the differences in this case of almost homogeneous fields remain significantly smaller than in the case of very strongly confined fields.

\begin{figure}[t]
    \centering
    \includegraphics{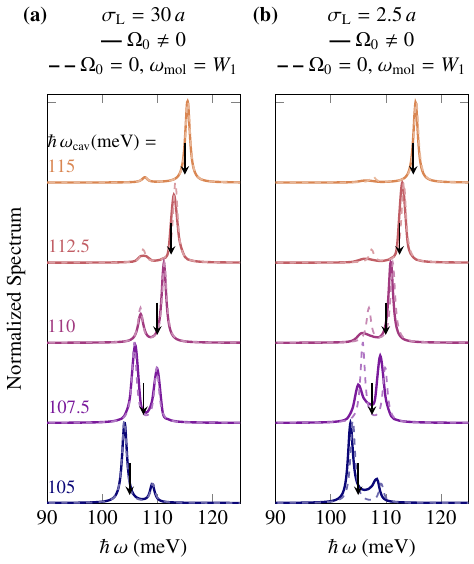}
    \caption{
    (Color online) Influence of direct molecule-molecule interactions for different cavity frequencies of an IR cavity characterized by \textbf{(a)} an almost homogeneous and \textbf{(b)} inhomogeneous field distribution.  The cavity is coupled to a 2D square lattice with $N=51 \times 51 = 2601$ molecules and the normalized optical spectra is plotted  for different cavity frequencies, indicated in the figure (arrows), and including (solid lines) or ignoring (dashed lines) direct molecule-molecule interactions. The former considers vibrational energy $\hbar\, \omega_\text{mol}= 100$~meV and the latter $\hbar\, \omega_\text{mol} = \hbar\, W_1 \approx 108.2$~meV. The spectra are shifted vertically for visibility and are obtained for a field distribution of width \textbf{(a)} $\sigma_\text{L} = 30\, a$ and \textbf{(b)} $\sigma_\text{L} = 2.5\, a$. Other parameters are dipole-dipole coupling strength $\hbar\, \Omega_{0} = 1$~meV, lattice constant $a=$ 0.5~nm, total molecules-cavity coupling strength $\hbar \, g_\text{tot} = \hbar \, \sqrt{\sum_j \md{g_j (\mathbf{r})}^2}=2$~meV and $\hbar\, \gamma = 1$~meV. 
    \label{Fig:Detuning}}
\end{figure}

We confirm the importance of direct molecule-molecule interactions for strong nanocavity field confinement by showing in Fig.~\ref{Fig:Detuning} the optical spectra as a function of the frequency of the cavity mode $\omega_\text{cav}$. We consider a small ($\sigma_\text{L}=30\,a$, Fig.~\ref{Fig:Detuning}~(a)) and a large ($\sigma_\text{L}=2.5\,a$, Fig.~\ref{Fig:Detuning}~(b)) field confinement, and other parameters are kept as as in  Fig.~\ref{Fig:Field}~(b) both when including (solid lines) and neglecting (dashed lines) the direct molecule-molecule interactions. In all cases, we find the typical avoided crossing near resonant conditions ($\hbar\, \omega_\text{cav} \approx \hbar\, W_1 \approx 108.2$~meV). Further, for weak field confinement, we find very similar results independently of the inclusion or not of the direct molecule-molecule interaction. In contrast, for the strongly confined nanocavity field, including the direct interaction (solid line in Fig.~\ref{Fig:Detuning}~(b)) strongly affects the results: by comparison, the peak spectrally closer to $\omega_\text{mol}=W_1$ becomes weaker and broader than the one near $\omega_\text{cav}$ (or the ones with no direct-direct interaction).

\subsection*{Establishing the criteria for strong dipole-dipole interaction}

We focus next on determining the condition that needs to be fulfilled for the direct molecule-molecule interactions to change significantly the optical response of a coupled system beyond a simple energy renormalization. With this purpose, we first consider that the cavity field mostly extends over a range $\approx \pr{0, k_\text{max}}$ of wavevectors in $k$-space. This range of wavevectors would indicate the set of modes from the dispersion (Fig.~\ref{Fig:2DWjN}~(e)) of the 2D material that can be excited. The numerical dispersion for small $\mathbf{k}$ can be approximated by (see derivation in App.~\ref{App1})

\begin{align}
    \omega \pc{\mathbf{k}} &= \sqrt{\omega^2_\text{mol} + 2 \omega_\text{mol} \, \sum_{m,\, n} \frac{\Omega_{0}}{\pc{\sqrt{m^2 + n^2}}^3} }  - 2 \pi \, \Omega_0 \md{\mathbf{k}} a,
    \label{eq:finaldispersionmaintext}
\end{align}
(the sum runs over $m, n = - N/2$ to $N/2$, excluding $n = m$),
so that the $\approx \pr{0, \md{\mathbf{k}_\text{max}}}$ range of wavevectors corresponds to frequencies covering a spectral width  $ \Delta\omega=\omega_\text{max} - \omega_\text{min} = 2 \pi \,\Omega_0 \md{\mathbf{k}_\text{max}} a$. We propose that the molecule-molecule interactions need to be considered explicitly in the Hamiltonian when this spectral width is of the order of, or larger than, the losses, i.e., $\Delta\omega\gtrapprox \gamma$, corresponding to the following condition:
\begin{align}
    2 \pi \, \Omega_0 \frac{a}{\sigma_\text{L}} \gtrapprox \gamma.
    \label{eq:condition}
\end{align}
We have approximated $\md{\mathbf{k}_\text{max} }\approx 1/ \sigma_\text{L}$ for the Gaussian illumination of width (in real space) $\sigma_\text{L}$. Equation~\eqref{eq:condition} predicts that the direct molecule-molecule interaction needs to be included in the Hamiltonian for very large coupling strength, very large confinement (low $\sigma_\text{L}$) and/or very low losses. This equation could be further generalized  to other conditions, such as different illuminations. This criteria depends only on the ratio of the different parameters, $\sigma_L/a$ and $\Omega_0/\gamma$, not on their absolute value.

\begin{figure}[t!]
    \centering
    \includegraphics{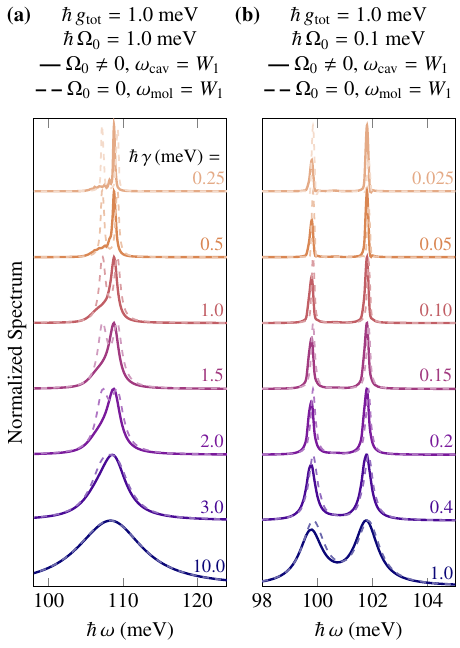}
    \caption{
    (Color online)     
    Influence of losses $\gamma$ in the normalized spectra. The cavity is coupled to a 2D square lattice with $N=51 \times 51 = 2601$ molecules and the normalized optical spectra is plotted for different cavity losses, $\gamma$, indicated in the figure, and including (solid lines) or ignoring (dashed lines) direct molecule-molecule interactions. The former considers vibrational energy $\hbar\, \omega_\text{mol}= 100$~meV and the latter $\omega_\text{mol} = W_1$. The spectra are shifted vertically for visibility and are obtained for a field distribution of width $\sigma_\text{L} = 2.5\, a$. Other parameters are dipole-dipole coupling strength \textbf{(a)} $\hbar\, \Omega_{0} = 1~\text{meV}$ $\pc{\hbar \, W_1 \approx 108.2~\text{meV}}$, and \textbf{(b)} $\hbar\, \Omega_{0} = 0.1~\text{meV}$ $\pc{\hbar \, W_1 \approx 100.8~\text{meV}}$, as well as $\omega_\text{cav} = W_1$, lattice constant $a=$ 0.5~nm and total molecules-cavity coupling strength $\hbar \, g_\text{tot} = \hbar \, \sqrt{\sum_j \md{g_j (\mathbf{r})}^2}=1$~meV.
    \label{Fig:Conditions}}
\end{figure}

To assess the validity of the proposed equation, we first consider a particular example of dipole-dipole coupling strength, $\hbar\, \Omega_0 = 1$~meV, $\sigma_\text{L} = 2.5\,a$ and optical coupling strength $\hbar \, g_{\text{tot}} = 1$~meV. Figure~\ref{Fig:Conditions} shows the corresponding spectra as the losses varies $\gamma$, again including the dipole-dipole interactions in the Hamiltonian and setting $\hbar\, \omega_\text{mol}= 100$~meV (solid line), or by ignoring them and renormalizing the vibrational energy to $\omega_\text{mol}= W_1$ (dashed line). We observe that the two spectra start to differ for  $\hbar \, \gamma\lessapprox 2-3$ meV, consistent with the condition  $\hbar\, \gamma  \lessapprox 2.4$~meV obtained from eq.~\eqref{eq:condition}. Interestingly, the spectra obtained for the weakest losses ($\hbar \, \gamma=0.24$~meV) and including dipole-dipole interactions shows many small narrow peaks, a direct signature of the participation of more than one collective vibrational mode in the response. 

As the threshold condition of $\gamma$ scales directly with $\Omega_0$, we next perform in Fig.~\ref{Fig:Conditions}~(b) a similar analysis for a much smaller value of the dipole-dipole interactions, $\hbar\, \Omega_0 = 0.1$~meV. This panel corresponds to numerical parameters closer to that of molecules. For this case, we observe that the two spectra start to differ for $\hbar\, \gamma  \lessapprox 0.2 - 0.4$~meV, and the condition in eq.~\eqref{eq:condition} gives $\hbar\, \gamma  \lessapprox 0.24$~meV, challenging to achieve experimentally but within reach \cite{JChemPhys156.104301(2022)}.
Last, we apply the criterion in eq.~\eqref{eq:condition} to the results in Fig.~\ref{Fig:Field}~(b). In this case, $\hbar\, \Omega_0 = \hbar\gamma=1$~meV , which gives the threshold $\sigma_\text{L} /a \lessapprox 2 \pi$ when the dipole-dipole interactions need to be explicitly considered, in good agreement with the numerical results.

%

\section{Conclusions \label{sec:conclusions}}

We have studied the effect of direct molecular dipole-dipole interactions on the optical response of photonic nanocavities strongly coupled with molecular assemblies or 2D materials. The description is based on a c-QED approach to the dynamics of the states in the system, without rotating wave approximation and including the losses via an effective broadening of the modes. An alternative approach to model the system based on classical dipoles or coupled harmonic oscillators is also possible.

As a first step, we describe how, in the absence of the nanocavity, the dipole-dipole interactions lead to the emergence of collective modes that span a significant frequency range and are characterized by a very large wavevector in the in-plane direction. It is thus possible to obtain a dispersion relation, and we find that it resembles the one characterizing an infinite layer, even for a relatively small molecular patch with a lateral size of only $N\cdot a \approx 25$\ nm (to be compared with the much larger wavelength).

When the photonic nanocavity is included, the influence of the direct molecule-molecule interactions on the optical response of the coupled system strongly depends on the degree of localization of the cavity fields. For homogeneous or slowly varying fields, the effect of these interactions is mostly a direct renormalization of the vibrational resonant frequency. However, when the fields are confined to the level of a few intermolecular distances, there exist conditions under which direct interactions can show a more profound effect. Instead of observing the  two clear polaritonic peaks, as it occurs for no direct interactions, one of the polaritonic modes can spread out into many (weaker) subpeaks. We attribute this observation to the fact that, as a consequence of the direct molecule-molecule interaction, multiple collective modes at different frequencies and wavevectors can couple (less efficiently) with the photonic nanocavity and contribute to the response.  

We have further derived a simple equation that indicates the conditions for this effect of dipole-dipole interaction to become relevant, beyond the energy renormalization. 
In general, molecules (or 2D materials) characterized by strong transition dipole moment and/or weak losses are required, together with large field localization. Several 2D materials are characterized by low losses and large vibrational dipole moments, which could reach the dipole-dipole coupling strength $\hbar \, \Omega_0\approx 1$~meV used in most of our calculations (App.~\ref{App:Mol2Dmat}). Molecules are generally characterized by weaker $\Omega_0$, but their associated losses could  also be sufficiently small. 
For example, several vibrations of the molecule 4,4'-bis(Ncarbazolyl)-1,1'-biphenyl (CBP) have losses of $\gamma_{mol} \approx 0.7-1.74$~meV $\approx 0.04-0.1~ \omega_\text{mol}$  and the losses of h-BN lattice vibrations (phonons) are $\gamma_\text{hBN} \approx 0.5-0.6$~meV \cite{JChemPhys156.104301(2022)}. Further, the vibrational losses for a single molecule embedded in a matrix can be as small as $0.07$~meV ($10$~ps lifetime).

Our results thus identify the conditions where it becomes necessary to include explicitly direct molecule-molecule interactions to describe the optical response of a coupled system beyond a renormalization of the energy.
Furthermore, we have focused our study on the coupling with molecular vibrations or phonons, but the analysis and conclusions can be applied to other transitions (such as molecular excitonic transitions) in a straightforward manner. For example, it is interesting to consider the coupling of plasmonic systems with excitonic transitions characterized by large dipole moment and present in quantum emitters such as molecules or quantum dots. The coupling strength can be very large in these systems \cite{Nature535.127(2016), Santhosh2016, Zhang2017}, and the losses of the excitonic transitions are only limited by the spontaneous decay rate and can thus be in principle extremely small at cryogenic temperature. On the other hand, plasmonic losses and room-temperature excitonic losses can be large, so that, it is important to consider the details of each experimental configuration.

\section*{Acknowledgments}
We would like to acknowledge Unai Muniain for fruitful discussions. This work was supported by Grant No. PID2019-107432GB-I00 funded by MCIN/AEI/10.13039/5 and Grant No. IT 1526-22 from the Basque Government for consolidated groups of the Basque University, as well as the European Union (NextGenerationEU) through the Complementary Plans (Grant No. PRTR-C17.I1) promoted by the Ministry of Science and Innovation within the Recovery, Transformation, and Resilience Plan of Spain, and is part of the activities of the IKUR strategy of the Department of Education promoted by the Basque Government.

\appendix
\section{Estimation of $\Omega_0$ for real materials \label{App:Mol2Dmat}}

To estimate the value of $\Omega_0$, we consider first a simple description of the relative permittivity $\varepsilon_{r}$ of a polar material at a frequency $\omega$, near a phonon resonance. Ignoring losses,
\begin{equation}
\varepsilon_r (\omega)=\varepsilon_\infty\left(1+\frac{\omega^2_L-\omega^2_T}{\omega_T^2-\omega^2} \right),
\end{equation}
where $\omega_L$ and $\omega_T$ are the longitudinal and transverse phonon frequency, respectively, and $\varepsilon_\infty$ the high-frequency permittivity. This permittivity corresponds to a transition dipole moment $d_{u}$ per unit cell of volume $V_\text{cell}$ \cite{NatComm12.6206(2021)}
\begin{equation}
d_{u}=\sqrt{\frac{\hbar}{2\omega_{T}}V_\text{cell}\varepsilon_0\varepsilon_\infty(\omega^2_L-\omega^2_T)},
\end{equation}
where $\varepsilon_0$ is the vacuum permittivity.
Assuming, for simplicity, a cubic structure, $V_\text{cell}=a^3$, with $a$ the lattice parameter of the unit cell, we obtain
\begin{equation}
\Omega_0=\frac{\md{d_u}^2}{4\pi\varepsilon_0\varepsilon_\infty \hbar a^3}=\frac{\omega^2_L-\omega^2_T}{8\pi\omega_T}
\end{equation}
For this derivation, we have considered an arbitrary value of $\varepsilon_\infty$ to show that it does not affect the final result. The model in the main text corresponds to $\varepsilon_\infty=1$ as only one vibrational mode is included. 

As an example, we obtain $\Omega_0=0.0196 \, \omega_T$ for SiC \cite{PhysRevB.55.10105(1997)}, and $\Omega_0=0.016\, \omega_T$ and $\Omega_0=0.008 \, \omega_T$ for the in plane and out of plane mode of hBN, respectively \cite{NatComm12.6206(2021)}. These values are comparable to the value  $\Omega_0=0.01\, \omega_\text{mol}$ used in the main text, although a more rigorous model would need to consider that the unit cell of these materials is not cubic, that we are comparing here with a bulk material and not a monolayer, and that $\omega_\text{mol}$ does not exactly correspond to $\omega_L$ or $\omega_T$.

We can also proceed in the same manner for molecular ensembles. We note that the permittivity is in this case typically written in a different form, such as
\begin{equation}
\varepsilon_r (\omega) =\varepsilon_\infty+\frac{S^2}{\omega_\text{mol}^2-\omega^2},
\end{equation}
where $S$ sets the strength of the resonance and $\omega_\text{mol}$ is the resonant frequency. Proceeding in the same way as before, we obtain 
\begin{equation}
\Omega_0=\frac{S^2}{8\pi\omega_{mol}\varepsilon_\infty}.
\end{equation}
For 4,4'-bis(N-carbazolyl)-1,1'-biphenyl (CBP) \cite{LightSciAppl7.17172(2018)}, $\varepsilon_\infty=2.8$ and the vibrational peak at $\omega_\text{mol}=1504$~cm$^{-1}$ is characterized by $S=164$~cm$^{-1}$, which gives $\Omega_0=1.7\times10^{-4}\omega_\text{mol}$.

\section{Optical response of the vibrational modes with lattice constant $a$ \label{App:Extra}}
The lattice constant will change the results of our discussion in the main text  quantitatively, but not qualitatively. For homogeneous fields, changing the lattice constant would have the same effect as changing the coupling strength, $\Omega_0\propto 1/a^3$ which we explore in Fig.~\ref{Fig:FitCav}~(b). For inhomogeneous fields, i.e., confined fields, changing $a$ can involve several effects simultaneously (change of $\Omega_0$, change of the normalized confinement $\sigma_L/a$, change of the coupling strength $g_j$).

\begin{figure}[t!]
    \centering
    \includegraphics{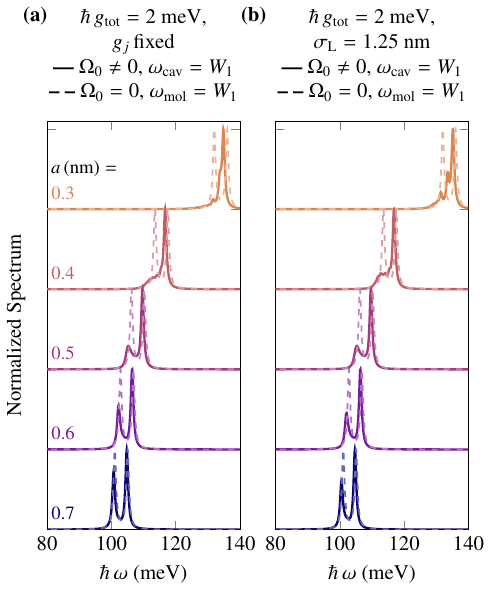}
    \caption{(Color online) Influence of the direct dipole-dipole interactions on the optical spectra $S (\omega)$ for different values of the lattice constant $a$. The normalized optical spectra of an IR cavity coupled to a 2D square lattice with $N=51 \times 51 = 2601$ molecules is plotted, keeping the total coupling fixed at $\hbar \, g_\text{tot} =2$~meV and including (solid lines) or ignoring (dashed lines) direct molecule-molecule interactions. The former considers vibrational energy $\hbar \, \omega_\text{mol} = 100$~meV and the latter $\omega_\text{mol} =W_1$. \textbf{(a)} Each equivalent molecule (in the same lattice position $j$) experiences the same coupling strength $g_j$ which requires to change $\sigma_\text{L}$ appropriately (for reference, when $a=0.5$~nm, $\sigma_\text{L} = 2.5\,a$). \textbf{(b)} Same as before, but the field distribution $\sigma_\text{L} = 1.25$~nm is kept fixed.
    The spectra are shifted vertically for visibility and are obtained for $\hbar \, \gamma = 1$~meV, and $\hbar\, \Omega_{0} = (0.125/a^3)~\text{meV}$. In all the spectra, the value of the resonant cavity frequency $\omega_\text{cav}$ is set to be equal to the corresponding value of $W_1$ .
    \label{Fig:LatticeChange}}
\end{figure}

Here, we explore two different situations where we change the lattice constant $a$. In Fig.~\ref{Fig:LatticeChange}~(a), we show the case where we fix $g_j$ and $g_\text{tot}$, and in Fig.~\ref{Fig:LatticeChange}~(b) the case where we fix $g_\text{tot}$ and $\sigma_\text{L}$. We show the results for a nanophotonic cavity which couples with the 2D molecular square patch containing $N=51 \times 51 = 2601$ molecules as a function of the lattice constant $a$, considering $\hbar \, \Omega_0 = (0.125/a^3)$~meV. We set for each case $\omega_\text{cav}= W_1$, and $\hbar \, g_\text{tot} = 2$~meV.
%

\section{Theoretical dispersion relation of an infinite monolayer \label{App1}}
In the main text, we have numerically solved for the collective excitations of the finite system arising from the direct dipole-dipole interactions. On the other hand, it is possible to obtain the analytical dispersion relation of the infinite 2D molecular monolayer by taking a solid-state approach, where quasi-excitations propagate on a lattice.  Each molecule is indexed by $\mathbf{s}$ and is located at $\mathbf{r}_s$. The total vibronic Hamiltonian becomes
\begin{align}
    \hH_\text{mol} + \hH_\text{vib-vib} &=  \hbar \sum_\mathbf{s} \omega_\text{mol} \hb^\dagger_\mathbf{s} \hb_\mathbf{s} \nonumber \\
    &\quad
    +  \hbar \sum_\mathbf{s} \sum_\mathbf{s' > s} \Omega_{
    \md{\mathbf{r}_\mathbf{s'}-\mathbf{r}_\mathbf{s}} }
    \pc{\hb^\dagger_{ \mathbf{s}} + \hb_{\mathbf{s}}}
\pc{\hb^\dagger_{\mathbf{s'} } + \hb_{\mathbf{s'}}}.
\end{align}
The equations of motion for the expectation values $\beta_{\mathbf{s}} = \av{\hb_{\mathbf{s}}}$  can be obtained from $d \hb_{\mathbf{s}} / dt = -i / \hbar \pr{\hb_{\mathbf{s}}, \hH_\text{mol} + \hH_\text{vib-vib}}$. To better illustrate the procedure to obtain the dispersion relation, we first assume nearest neighbor coupling $\Omega_{
    \md{\mathbf{r}_\mathbf{s'}-\mathbf{r}_\mathbf{s}} } = \Omega_{j,j\pm 1} \equiv \Omega_0$ and neglect the terms $\hb_{ \mathbf{s}}\hb_{ \mathbf{s'}}$ and $\hb^\dagger_{ \mathbf{s}}\hb^\dagger_{ \mathbf{s'}}$ that describe the creation and annihilation of two vibrations at the same time (rotating wave approximation, RWA).
By doing so, we arrive at
\begin{align*}
    \dot{\beta}_j
    &= - i \omega_\text{mol}\, \beta_j - i \, \Omega_0 \pc{ \beta_{j-1} + \beta_{j+1} }.
\end{align*}
We then insert the ansatz $\beta_j = \mathcal{B}_k e^{i(k a j - \omega t)}$ into the above equation, where $a$ is the lattice constant, and obtain
\begin{align}
    \omega &= \omega_\text{mol} + \Omega_0 \pc{e^{ik a} + e^{-ik a}},
\end{align}
which straightforwardly leads to the following dispersion relation
\begin{align}
    \omega = \omega_\text{mol} + 2 \Omega_0 \cos (k a).
\end{align}

This result has the advantage of simplicity, but does not show the right tendency for low values of $\mathbf{k}$, where the coupling with a significant number of neighbors beyond the nearest one can contribute to the result. However,
the procedure described can be extended to all neighbors in a straightforward manner. The equations of motions are in this case $\dot{\beta}_j = - i \omega_\text{mol}\, \beta_j - i \sum_{l \neq j} \Omega_{jl} \, \beta_{l}$. Using the ansatz $\beta_j = \mathcal{B}_k e^{i(\mathbf{k}\cdot \mathbf{r}_{lj} - \omega t)}$ with $\mathbf{r}_{lj} = \mathbf{r}_l - \mathbf{r}_j$ gives the dispersion relation
\begin{align}
    \omega = \omega_0 + \sum_{l \neq j} \frac{\Omega_{lj}}{2} \,\pc{ e^{i\, \mathbf{k}\cdot \mathbf{r}_{lj}} + e^{-i\, \mathbf{k}\cdot \mathbf{r}_{lj}} },
    \label{eq:dispersionRWA}
\end{align}
with the factor of $1/2$ to avoid double counting, $\Omega_{lj}$ as defined in eq.~\eqref{eq:p0}, and the sum runs over all $l \neq j$, with $j$ an arbitrary number

\subsection{Dispersion relation beyond RWA approximation}

For the relatively small values of $\Omega_0$ used in the main text, the RWA (used to derive Eq.~\eqref{eq:dispersionRWA}) is a good approximation. However, for completeness, we derive next the dispersion relation without the RWA, i.e. including  the terms that do not conserve the number of excitations. We begin by considering again only coupling between nearest neighbors in a 1D chain. 
The equations of motion become
\begin{align*}
    \dot{\beta}_j &= - i \omega_\text{mol} \, \beta_j - i \, \Omega_0 \, \beta_{j-1} - i\,  \Omega_0 \, \beta_{j+1} - i \, \Omega_0 \, \beta^*_{j-1} - i \, \Omega_0 \, \beta^*_{j+1} ,\\
    \dot{\beta}^*_j &= i \omega_\text{mol} \, \beta^*_j +i \, \Omega_0 \, \beta^*_{j-1} + i \, \Omega_0 \,  \beta^*_{j+1} + i \, \Omega_0 \, \beta_{j-1} + i \, \Omega_0 \beta_{j+1}.
\end{align*}
Defining $A_j = \beta_j + \beta^*_j$ and $B_j = \beta_j - \beta^*_j$, one gets
\begin{align*}
    \frac{d}{dt} A_j &= - i \omega_\text{mol} \, B_j, \\
    \frac{d}{dt} B_j  &= -i \omega_\text{mol} \, A_j - i 2\Omega_0 \pc{ A_{j-1} + A_{j+1}}.
\end{align*}
Taking the time derivative of the first equation, we find
\begin{align*}
    \frac{d^2}{dt^2} A_j &=  - \omega^2_j \, A_j - 2 \omega_\text{mol} \Omega_0 \pc{ A_{j-1}  + A_{j+1}}.
\end{align*}
Following the same procedure as before, with the ansatz $A_j = e^{i(k a j - \omega t)}$, results in
\begin{align}
     \omega &=  \sqrt{\omega^2_j + 2 \omega_\text{mol} \, \Omega_0 \pc{ e^{i k a } + e^{-i k a}} }
\end{align}
On the other hand, if we proceed in the same manner but expanding to all possible neighbors, we obtain the final dispersion relation
\begin{align}
     \omega &=  \sqrt{\omega^2_\text{mol} + \omega_\text{mol} \sum_{l \neq  j} \Omega_{lj} \,\pc{ e^{i\, \mathbf{k}\cdot \mathbf{r}_{lj}} + e^{-i\, \mathbf{k}\cdot \mathbf{r}_{lj}} }},
    \label{eq:dispersionnoRWA}
\end{align}
with a factor $1/2$ again included when we expand to more molecules. 
Eq.~\eqref{eq:dispersionnoRWA} reduces to the RWA in the limit of small $\Omega_{lj}$, as expected.
\begin{figure}[t!]
    \centering
    \includegraphics{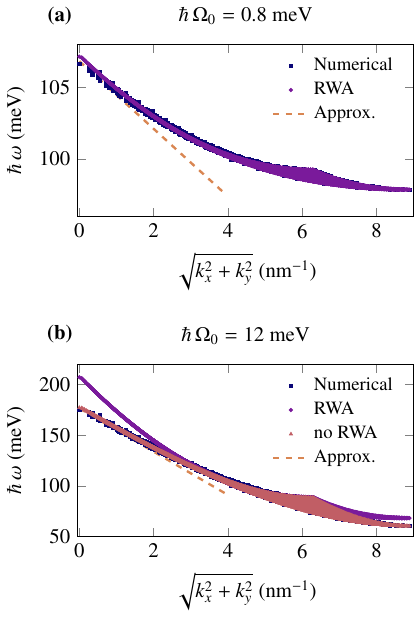}
    \caption{(Color online) Dispersion relation for dipole-dipole interaction strength between the molecules, \textbf{(a)} $\hbar\, \Omega_{0} = 0.8$ and \textbf{(b)} 12~meV. The molecules are placed in a square lattice with lattice constant $a=0.5$~nm, and $\hbar\, \omega_\text{mol} = 100~\text{meV}$. We compare the numerical results of the eigenvalues
    $W_n$ for  $N=51 \times 51 = 2601$ molecules (blue squares) with the analytical expression for an infinite monolayer obtained with (lilac circles, eq.~\eqref{eq:dispersionRWA}) and without (pink triangles, eq.~\eqref{eq:dispersionnoRWA}) the RWA. 
    The analytical equations are evaluated for a range of $k_{x,\,y} =\pr{0,\, \pi/a}$. In dashed lines, we show the linear approximation according to eq.~\eqref{eq:finaldispersionappendix}.}
    \label{fig:disprelcomp}
\end{figure}
\begin{figure}[t!]
    \centering
    \includegraphics{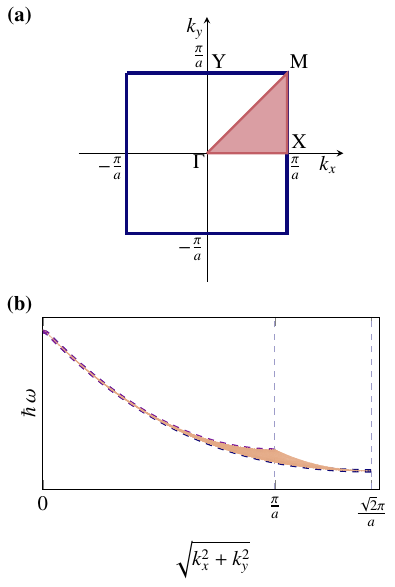}
    \caption{\textbf{(a)} Brillouin zone of the square lattice. The critical points (points of high symmetry) $\Gamma$, M, X and Y, are depicted. \textbf{(b)} An illustrative example of a typical dispersion relation for a square lattice with dipole-dipole interactions (shaded orange area). For a fixed $\sqrt{k_x^2 + k_y^2}$, the energies of the collective vibrational modes show a certain spread, so that, they are contained within the two dashed blue lines shown in the plot. The upper dashed blue line corresponds to wavevectors in the $\Gamma X$ direction in the reciprocal space, with $k_y = 0 $ and $k_{x} \le \pi / a$. The lower dashed blue line corresponds to wavevectors in the  $\Gamma M$ direction ($ k_x = k_y$ and $\sqrt{\left(k_x^\text{max}\right)^2 + \left(k_y^\text{max}\right)^2} = \sqrt{2} (\pi / a)$. \label{Fig:Brillouin}}
\end{figure}

In Fig.~\ref{fig:disprelcomp}, we compare the numerical dispersion relation for a finite ($N = 51\times 51$) number of molecules (without nanocavity), obtained as in Fig.~\ref{Fig:2DWjN}~(e), with the analytical dispersion of the infinite monolayer. The latter is evaluated in the range $k_{x,\,y} =\pr{0,\, \pi/a}$ with increments of $\pi / (100\, a)$.
The results for $\hbar\, \Omega_0=0.8$~meV in Fig.~\ref{fig:disprelcomp}~(a) show good agreement between the numerical results of the finite system (blue squares) and the analytical results of the infinite monolayer within the RWA (eq.~\eqref{eq:dispersionRWA}, lilac circles). The exact theoretical dispersion of the infinite monolayer obtained without the RWA (eq.~\eqref{eq:dispersionnoRWA}) is not shown, but is almost identical to that obtained with the RWA. 

To illustrate a situation that requires going beyond the RWA, we plot in Fig.~\ref{fig:disprelcomp}~(b) the dispersion relation for a very large $\hbar\, \Omega_0=12$~meV. This large coupling leads to a significantly larger span of  energies of the collective modes. We find again that the exact theoretical dispersion of the infinite system, as obtained without RWA (orange triangles), eq.~\eqref{eq:dispersionnoRWA}, matches well the numerical results for the finite number of molecules (blue squares). However, the theoretical dispersion of the infinite system obtained within the RWA (lilac circles), eq.~\eqref{eq:dispersionRWA}, is markedly different from both.  

Fig.~\ref{fig:disprelcomp} also shows a `spreading' of the data points in the dispersion relation. To explain its origin, we plot in Fig.~\ref{Fig:Brillouin}~(a) the Brillouin zone for a lattice with a square unit cell, where we enhance the triangle formed by the critical points $\Gamma$, X, M. The dispersion is shown schematically in Fig.~\ref{Fig:Brillouin}~(b). For each wavevector $\md{\mathbf{k}}=\sqrt{k_x^2 + k_y^2}$, the eigenvalues extend over a range of energies. The eigenvectors with wavevector pointing in the  direction $\Gamma X$ of the Brillouin zone ($k_y = 0 $,  maximum value $k_{x}^\text{max}  = \pi / a$) correspond to the points of largest energy for each $\md{\mathbf{k}}$. Similarly,  eigenvectors with wavevector along the  $\Gamma M$ direction $\pc{k_x = k_y}$, with maximum value $\sqrt{\left(k_x^\text{max}\right)^2 + \left(k_y^\text{max}\right)^2} = \sqrt{2} (\pi / a)$, results in the lowest-energy solutions. The eigenenergies corresponding to these two directions are highlighted in the figure by the dashed blue lines.

\subsection{Linear approximation for $\mathbf{k} \to 0$}

The dispersion relations  in  Fig.~\ref{fig:disprelcomp} show a linear dependence between the energy and the modulus $\md{\mathbf{k}}$ of the wavevector for small $\mathbf{k}$. For simplicity, we consider the case $k_y = 0$, as the results show the same slope with $\md{\mathbf{k}}=\sqrt{k_x^2 + k_y^2}$, independently of the individual weights of $k_x$ and $k_y$. Moreover, since we are considering a square lattice with constant $a$, let us write  the positions $x_l = a\, m$ and $y_j = a \, n$.
In this case, eq.~\eqref{eq:dispersionnoRWA} becomes 
\begin{align*}
     \omega &= \sqrt{\omega^2_\text{mol} + 2 \omega_\text{mol} \, \Omega_0 \sum_{m,\,n} \frac{\cos \pc{ \, k_x\,a\, m}}{\pc{\sqrt{m^2 + n^2}}^3}  },
\end{align*}
where the sum runs over $m, n = - N/2$ to $N/2$, excluding $n = m$.
At $k_x=0$, this expression takes the value,
\begin{align*}
     \omega \pc{k_x = 0} &= \sqrt{\omega^2_\text{mol} + 2 \omega_\text{mol} \, \sum_{m,\, n} \frac{\Omega_{0}}{\pc{\sqrt{m^2 + n^2}}^3}} .
\end{align*}
To find the slope, we first focus on the term 
\begin{align*}
2 \omega_\text{mol} \, \Omega_0 \sum_{m ,\  n} \frac{ \cos \pc{ \, k_x\,a\, m} }{ \pc{\sqrt{m^2 + n^2}}^3}
\end{align*}
inside the square root. The derivative of this term with respect to $k_x$ is 
\begin{align*}
    - 2 \omega_\text{mol} \, \Omega_0 \sum_{m,\,  n} \frac{ a\, m \sin \pc{ \, k_x\,a\, m}}{\pc{\sqrt{m^2 + n^2}}^3}.
\end{align*}
Interestingly, all the individual terms are null at $k_x=0$, but the infinite sum is not, which indicates that the value of the sum at $k_x=0$ is determined by the terms corresponding to large $m^2+n^2$. Based on this, we convert the sum in $m$ and $n$ to an integral, which can be solved analytically:
\begin{align*}
    - 2 \omega_\text{mol} \, \Omega_0 \int^\infty_{-\infty} \!\!\!\! dx \int^\infty_{-\infty} \!\!\!\! dy \, \frac{ a x \sin \pc{ \, k_x \, a x}}{\pc{\sqrt{x^2 + y^2}}^3} = - 4 \pi \omega_\text{mol} \, \Omega_0 a.
\end{align*}

 The energies at low values of $k_x$ thus follows 
\begin{align*}
     \omega &= \sqrt{\omega^2_\text{mol}+ 2 \omega_\text{mol} \, \sum_{m,\, n} \frac{\Omega_{0}}{\pc{\sqrt{m^2 + n^2}}^3}  - 4 \pi \omega_\text{mol} \, \Omega_0\, a \,k_x }.
\end{align*}
Expanding this expression to first order and substituting $k_x\rightarrow \md{\mathbf{k}}=\sqrt{k_x^2 + k_y^2}$ (see above):
\begin{align}
     \omega &= \sqrt{\omega^2_\text{mol} + 2 \omega_\text{mol} \, \sum_{m,\, n} \frac{\Omega_{0}}{\pc{\sqrt{m^2 + n^2}}^3} } \nonumber \\
     &\quad - \frac{2 \pi\, \Omega_0 \md{\mathbf{k}} a \, \omega_\text{mol}}{\sqrt{\omega^2_\text{mol} + 2 \omega_\text{mol} \, \sum_{m,\, n} \frac{\Omega_{0}}{\pc{\sqrt{m^2 + n^2}}^3} }}.
     \label{eq:finaldispersionappendix}
\end{align}
For $\Omega_0\ll \omega_\text{mol}$ the second term of the denominator can be ignored and eq.~\eqref{eq:finaldispersionappendix} becomes eq.~\eqref{eq:finaldispersionmaintext} in the main text. The dispersion given by eq.~\eqref{eq:finaldispersionappendix} is also plotted in Fig.~\ref{fig:disprelcomp} (dashed line) and shows a good agreement with the numerical results for low $\md{\mathbf{k}}$.

\section{Comparing different ways of including losses \label{App:Losses}}
\begin{figure}
    \centering
    \includegraphics{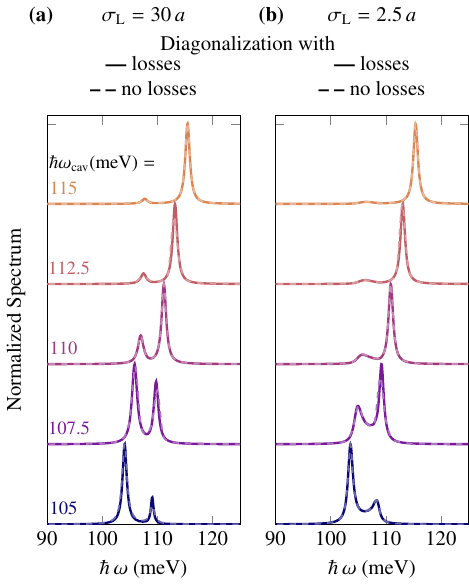}
    \caption{
    (Color online) Comparison of the results obtained using two different procedures to incorporate losses in the spectral response. Normalized spectra are shown for a 2D lattice with $N=51 \times 51 = 2601$ molecules with lattice constant 0.5~nm and $\hbar\, \omega_\text{mol} = 100~\text{meV}$ that couple with an IR cavity mode characterized by field confinement \textbf{(a)} $\sigma_\text{L} = 30\, a$ and \textbf{(b)} $\sigma_\text{L} = 2.5\, a$. In both cases, $\hbar \, g_\text{tot} = \hbar \, \sqrt{\sum_j \md{g_j (\mathbf{r})}^2} =2$~meV and the dipole-dipole coupling strength is $\hbar\, \Omega_{0} = 1 ~\text{meV}$ $\pc{\hbar \, W_1 \approx 108.2~\text{meV}}$. The solid lines correspond to the results obtained when incorporating  molecular, $\hbar\, \Gamma =1~\text{meV}$, and plasmonic losses, $\hbar \, \kappa = 1$~meV, directly in the Hamiltonian and applying eq.~\eqref{eq:SwLosses}. The dashed lines are obtained, as in the main text, by solving the Hamiltonian without losses and then applying eq.~\eqref{eq:SwNoInt} with line broadening, $\hbar\,  \gamma = 1$~meV. The spectra are shifted vertically for visibility.
    \label{Fig:DetuningLosses}}
\end{figure}

We incorporated the losses in the spectra of the main text by changing the delta functions obtained in the spectral function when no losses are considered into Lorentzian lines of full width at half maximum $\gamma$. In the following, we compare these results to those obtained by performing the transformation $\omega_\text{cav} \to \omega_\text{cav} + i \, \kappa / 2$ and $\omega_\text{mol} \to \omega_\text{mol} + i \, \Gamma/2 $ and numerically diagonalizing the full Hamiltonian of the system to find the new eigenvalues. For simplicity, we considered molecular losses $\hbar\, \Gamma = 1~\text{meV}$ and different cavities with $\hbar\, \kappa = 1~\text{meV}$. 

As the new eigenvectors of the system $\mathcal{W}_m$ are no longer real, we use the same definition as in eq.~\eqref{eq:SwNoInt}, but modified as
\begin{align}
    S(\omega) &= \sum_m \pc{\md{\zeta_{m1}}^2 -\md{\eta_{m1}}^2} \nonumber \\
    &\quad \times \frac{\Imag \pr{\mathcal{W}_m} }{(\omega-\Real \pr{\mathcal{W}_m})^2 + (\Imag \pr{\mathcal{W}_m} )^2},
    \label{eq:SwLosses}
\end{align}
in order to compare both results.

In Fig.~\ref{Fig:DetuningLosses}, we compare the results obtained in the main text (Eq.~\eqref{eq:SwNoInt}, dashed lines) with those obtained using the procedure described in this appendix(solid lines). The results are obtained for the same system as in Fig.~\ref{Fig:Setup}~(a) ($N = 51\times51$ dipoles in a monolayer coupled to a cavity mode), for small $\sigma_\text{L} = 30\, a$ (Fig.~\ref{Fig:DetuningLosses}~(a)), and large $\sigma_\text{L} = 2.5\, a$ (Fig.~\ref{Fig:DetuningLosses}~(b)) localization of the cavity fields. The results are very similar in both of the procedures used to incorporate losses.

\bibliography{Bibliography}

\begin{thebibliography}{60}%
\makeatletter
\providecommand \@ifxundefined [1]{%
 \@ifx{#1\undefined}
}%
\providecommand \@ifnum [1]{%
 \ifnum #1\expandafter \@firstoftwo
 \else \expandafter \@secondoftwo
 \fi
}%
\providecommand \@ifx [1]{%
 \ifx #1\expandafter \@firstoftwo
 \else \expandafter \@secondoftwo
 \fi
}%
\providecommand \natexlab [1]{#1}%
\providecommand \enquote  [1]{``#1''}%
\providecommand \bibnamefont  [1]{#1}%
\providecommand \bibfnamefont [1]{#1}%
\providecommand \citenamefont [1]{#1}%
\providecommand \href@noop [0]{\@secondoftwo}%
\providecommand \href [0]{\begingroup \@sanitize@url \@href}%
\providecommand \@href[1]{\@@startlink{#1}\@@href}%
\providecommand \@@href[1]{\endgroup#1\@@endlink}%
\providecommand \@sanitize@url [0]{\catcode `\\12\catcode `\$12\catcode
  `\&12\catcode `\#12\catcode `\^12\catcode `\_12\catcode `\%12\relax}%
\providecommand \@@startlink[1]{}%
\providecommand \@@endlink[0]{}%
\providecommand \url  [0]{\begingroup\@sanitize@url \@url }%
\providecommand \@url [1]{\endgroup\@href {#1}{\urlprefix }}%
\providecommand \urlprefix  [0]{URL }%
\providecommand \Eprint [0]{\href }%
\providecommand \doibase [0]{https://doi.org/}%
\providecommand \selectlanguage [0]{\@gobble}%
\providecommand \bibinfo  [0]{\@secondoftwo}%
\providecommand \bibfield  [0]{\@secondoftwo}%
\providecommand \translation [1]{[#1]}%
\providecommand \BibitemOpen [0]{}%
\providecommand \bibitemStop [0]{}%
\providecommand \bibitemNoStop [0]{.\EOS\space}%
\providecommand \EOS [0]{\spacefactor3000\relax}%
\providecommand \BibitemShut  [1]{\csname bibitem#1\endcsname}%
\let\auto@bib@innerbib\@empty
\bibitem [{\citenamefont {Agranovich}\ \emph {et~al.}(2003)\citenamefont
  {Agranovich}, \citenamefont {Litinskaia},\ and\ \citenamefont
  {Lidzey}}]{PhysRevB.67.085311(2003)}%
  \BibitemOpen
  \bibfield  {author} {\bibinfo {author} {\bibfnamefont {V.~M.}\ \bibnamefont
  {Agranovich}}, \bibinfo {author} {\bibfnamefont {M.}~\bibnamefont
  {Litinskaia}},\ and\ \bibinfo {author} {\bibfnamefont {D.~G.}\ \bibnamefont
  {Lidzey}},\ }\bibfield  {title} {\bibinfo {title} {Cavity polaritons in
  microcavities containing disordered organic semiconductors},\ }\href
  {https://doi.org/10.1103/PhysRevB.67.085311} {\bibfield  {journal} {\bibinfo
  {journal} {Phys. Rev. B}\ }\textbf {\bibinfo {volume} {67}},\ \bibinfo
  {pages} {085311} (\bibinfo {year} {2003})}\BibitemShut {NoStop}%
\bibitem [{\citenamefont {Long}\ and\ \citenamefont
  {Simpkins}(2015)}]{ACSPhot1.130(2015)}%
  \BibitemOpen
  \bibfield  {author} {\bibinfo {author} {\bibfnamefont {J.~P.}\ \bibnamefont
  {Long}}\ and\ \bibinfo {author} {\bibfnamefont {B.~S.}\ \bibnamefont
  {Simpkins}},\ }\bibfield  {title} {\bibinfo {title} {Coherent coupling
  between a molecular vibration and fabry–perot optical cavity to give
  hybridized states in the strong coupling limit},\ }\href
  {https://doi.org/10.1021/ph5003347} {\bibfield  {journal} {\bibinfo
  {journal} {ACS Photonics}\ }\textbf {\bibinfo {volume} {2}},\ \bibinfo
  {pages} {130} (\bibinfo {year} {2015})}\BibitemShut {NoStop}%
\bibitem [{\citenamefont {Shalabney}\ \emph {et~al.}(2015)\citenamefont
  {Shalabney}, \citenamefont {George}, \citenamefont {Hutchison}, \citenamefont
  {Pupillo}, \citenamefont {Genet},\ and\ \citenamefont
  {Ebbesen}}]{NatComm6.5981(2015)}%
  \BibitemOpen
  \bibfield  {author} {\bibinfo {author} {\bibfnamefont {A.}~\bibnamefont
  {Shalabney}}, \bibinfo {author} {\bibfnamefont {J.}~\bibnamefont {George}},
  \bibinfo {author} {\bibfnamefont {J.}~\bibnamefont {Hutchison}}, \bibinfo
  {author} {\bibfnamefont {G.}~\bibnamefont {Pupillo}}, \bibinfo {author}
  {\bibfnamefont {C.}~\bibnamefont {Genet}},\ and\ \bibinfo {author}
  {\bibfnamefont {T.~W.}\ \bibnamefont {Ebbesen}},\ }\bibfield  {title}
  {\bibinfo {title} {Coherent coupling of molecular resonators with a
  microcavity mode},\ }\href {https://doi.org/10.1038/ncomms6981} {\bibfield
  {journal} {\bibinfo  {journal} {Nat. Comm.}\ }\textbf {\bibinfo {volume}
  {6}},\ \bibinfo {pages} {5981} (\bibinfo {year} {2015})}\BibitemShut
  {NoStop}%
\bibitem [{\citenamefont {George}\ \emph {et~al.}(2015)\citenamefont {George},
  \citenamefont {Shalabney}, \citenamefont {Hutchison}, \citenamefont {Genet},\
  and\ \citenamefont {Ebbesen}}]{JPhysChemLett6.1027(2015)}%
  \BibitemOpen
  \bibfield  {author} {\bibinfo {author} {\bibfnamefont {J.}~\bibnamefont
  {George}}, \bibinfo {author} {\bibfnamefont {A.}~\bibnamefont {Shalabney}},
  \bibinfo {author} {\bibfnamefont {J.~A.}\ \bibnamefont {Hutchison}}, \bibinfo
  {author} {\bibfnamefont {C.}~\bibnamefont {Genet}},\ and\ \bibinfo {author}
  {\bibfnamefont {T.~W.}\ \bibnamefont {Ebbesen}},\ }\bibfield  {title}
  {\bibinfo {title} {Liquid-phase vibrational strong coupling},\ }\href
  {https://doi.org/10.1021/acs.jpclett.5b00204} {\bibfield  {journal} {\bibinfo
   {journal} {J. Phys. Chem. Lett.}\ }\textbf {\bibinfo {volume} {6}},\
  \bibinfo {pages} {1027} (\bibinfo {year} {2015})}\BibitemShut {NoStop}%
\bibitem [{\citenamefont {Simpkins}\ \emph {et~al.}(2015)\citenamefont
  {Simpkins}, \citenamefont {Fears}, \citenamefont {Dressick}, \citenamefont
  {Spann}, \citenamefont {Dunkelberger},\ and\ \citenamefont
  {Owrutsky}}]{ACSPhot10.1460(2015)}%
  \BibitemOpen
  \bibfield  {author} {\bibinfo {author} {\bibfnamefont {B.~S.}\ \bibnamefont
  {Simpkins}}, \bibinfo {author} {\bibfnamefont {K.~P.}\ \bibnamefont {Fears}},
  \bibinfo {author} {\bibfnamefont {W.~J.}\ \bibnamefont {Dressick}}, \bibinfo
  {author} {\bibfnamefont {B.~T.}\ \bibnamefont {Spann}}, \bibinfo {author}
  {\bibfnamefont {A.~D.}\ \bibnamefont {Dunkelberger}},\ and\ \bibinfo {author}
  {\bibfnamefont {J.~C.}\ \bibnamefont {Owrutsky}},\ }\bibfield  {title}
  {\bibinfo {title} {Spanning strong to weak normal mode coupling between
  vibrational and fabry–pérot cavity modes through tuning of vibrational
  absorption strength},\ }\href {https://doi.org/10.1021/acsphotonics.5b00324}
  {\bibfield  {journal} {\bibinfo  {journal} {ACS Photonics}\ }\textbf
  {\bibinfo {volume} {2}},\ \bibinfo {pages} {1460} (\bibinfo {year}
  {2015})}\BibitemShut {NoStop}%
\bibitem [{\citenamefont {Lather}\ \emph {et~al.}(2019)\citenamefont {Lather},
  \citenamefont {Bhatt}, \citenamefont {Thomas}, \citenamefont {Ebbesen},\ and\
  \citenamefont {George}}]{AngChem58.10635(2019)}%
  \BibitemOpen
  \bibfield  {author} {\bibinfo {author} {\bibfnamefont {J.}~\bibnamefont
  {Lather}}, \bibinfo {author} {\bibfnamefont {P.}~\bibnamefont {Bhatt}},
  \bibinfo {author} {\bibfnamefont {A.}~\bibnamefont {Thomas}}, \bibinfo
  {author} {\bibfnamefont {T.~W.}\ \bibnamefont {Ebbesen}},\ and\ \bibinfo
  {author} {\bibfnamefont {J.}~\bibnamefont {George}},\ }\bibfield  {title}
  {\bibinfo {title} {Cavity catalysis by cooperative vibrational strong
  coupling of reactant and solvent molecules},\ }\href
  {https://doi.org/10.1002/anie.201905407} {\bibfield  {journal} {\bibinfo
  {journal} {Angew. Chem.}\ }\textbf {\bibinfo {volume} {58}},\ \bibinfo
  {pages} {10635} (\bibinfo {year} {2019})}\BibitemShut {NoStop}%
\bibitem [{\citenamefont {Lidzey}\ \emph {et~al.}(1998)\citenamefont {Lidzey},
  \citenamefont {Bradley}, \citenamefont {Skolnick}, \citenamefont {Virgili},
  \citenamefont {Walker},\ and\ \citenamefont
  {Whittaker}}]{Nature395.53(1998)}%
  \BibitemOpen
  \bibfield  {author} {\bibinfo {author} {\bibfnamefont {D.~G.}\ \bibnamefont
  {Lidzey}}, \bibinfo {author} {\bibfnamefont {D.~D.~C.}\ \bibnamefont
  {Bradley}}, \bibinfo {author} {\bibfnamefont {M.~S.}\ \bibnamefont
  {Skolnick}}, \bibinfo {author} {\bibfnamefont {T.}~\bibnamefont {Virgili}},
  \bibinfo {author} {\bibfnamefont {S.}~\bibnamefont {Walker}},\ and\ \bibinfo
  {author} {\bibfnamefont {D.~M.}\ \bibnamefont {Whittaker}},\ }\bibfield
  {title} {\bibinfo {title} {Strong exciton--photon coupling in an organic
  semiconductor microcavity},\ }\href {https://doi.org/10.1038/25692}
  {\bibfield  {journal} {\bibinfo  {journal} {Nature}\ }\textbf {\bibinfo
  {volume} {395}},\ \bibinfo {pages} {53} (\bibinfo {year} {1998})}\BibitemShut
  {NoStop}%
\bibitem [{\citenamefont {Lidzey}\ \emph {et~al.}(1999)\citenamefont {Lidzey},
  \citenamefont {Bradley}, \citenamefont {Virgili}, \citenamefont {Armitage},
  \citenamefont {Skolnick},\ and\ \citenamefont
  {Walker}}]{PhysRevLett82.3316(1999)}%
  \BibitemOpen
  \bibfield  {author} {\bibinfo {author} {\bibfnamefont {D.~G.}\ \bibnamefont
  {Lidzey}}, \bibinfo {author} {\bibfnamefont {D.~D.~C.}\ \bibnamefont
  {Bradley}}, \bibinfo {author} {\bibfnamefont {T.}~\bibnamefont {Virgili}},
  \bibinfo {author} {\bibfnamefont {A.}~\bibnamefont {Armitage}}, \bibinfo
  {author} {\bibfnamefont {M.~S.}\ \bibnamefont {Skolnick}},\ and\ \bibinfo
  {author} {\bibfnamefont {S.}~\bibnamefont {Walker}},\ }\bibfield  {title}
  {\bibinfo {title} {Room temperature polariton emission from strongly coupled
  organic semiconductor microcavities},\ }\href
  {https://doi.org/10.1103/PhysRevLett.82.3316} {\bibfield  {journal} {\bibinfo
   {journal} {Phys. Rev. Lett.}\ }\textbf {\bibinfo {volume} {82}},\ \bibinfo
  {pages} {3316} (\bibinfo {year} {1999})}\BibitemShut {NoStop}%
\bibitem [{\citenamefont {Lidzey}\ \emph {et~al.}(2000)\citenamefont {Lidzey},
  \citenamefont {Bradley}, \citenamefont {Armitage}, \citenamefont {Walker},\
  and\ \citenamefont {Skolnick}}]{Science288.1620(2000)}%
  \BibitemOpen
  \bibfield  {author} {\bibinfo {author} {\bibfnamefont {D.~G.}\ \bibnamefont
  {Lidzey}}, \bibinfo {author} {\bibfnamefont {D.~D.~C.}\ \bibnamefont
  {Bradley}}, \bibinfo {author} {\bibfnamefont {A.}~\bibnamefont {Armitage}},
  \bibinfo {author} {\bibfnamefont {S.}~\bibnamefont {Walker}},\ and\ \bibinfo
  {author} {\bibfnamefont {M.~S.}\ \bibnamefont {Skolnick}},\ }\bibfield
  {title} {\bibinfo {title} {Photon-mediated hybridization of frenkel excitons
  in organic semiconductor microcavities},\ }\href
  {https://doi.org/10.1126/science.288.5471.1620} {\bibfield  {journal}
  {\bibinfo  {journal} {Science}\ }\textbf {\bibinfo {volume} {288}},\ \bibinfo
  {pages} {1620} (\bibinfo {year} {2000})}\BibitemShut {NoStop}%
\bibitem [{\citenamefont {Leng}\ \emph {et~al.}(2018)\citenamefont {Leng},
  \citenamefont {Szychowski}, \citenamefont {Daniel},\ and\ \citenamefont
  {Pelton}}]{NatComm.9.4012(2018)}%
  \BibitemOpen
  \bibfield  {author} {\bibinfo {author} {\bibfnamefont {H.}~\bibnamefont
  {Leng}}, \bibinfo {author} {\bibfnamefont {B.}~\bibnamefont {Szychowski}},
  \bibinfo {author} {\bibfnamefont {M.-C.}\ \bibnamefont {Daniel}},\ and\
  \bibinfo {author} {\bibfnamefont {M.}~\bibnamefont {Pelton}},\ }\bibfield
  {title} {\bibinfo {title} {Strong coupling and induced transparency at room
  temperature with single quantum dots and gap plasmons},\ }\href
  {https://doi.org/10.1038/s41467-018-06450-4} {\bibfield  {journal} {\bibinfo
  {journal} {Nat. Com.}\ }\textbf {\bibinfo {volume} {9}},\ \bibinfo {pages}
  {4012} (\bibinfo {year} {2018})}\BibitemShut {NoStop}%
\bibitem [{\citenamefont {Schachenmayer}\ \emph {et~al.}(2015)\citenamefont
  {Schachenmayer}, \citenamefont {Genes}, \citenamefont {Tignone},\ and\
  \citenamefont {Pupillo}}]{PhysRevLett.114.196403(2015)}%
  \BibitemOpen
  \bibfield  {author} {\bibinfo {author} {\bibfnamefont {J.}~\bibnamefont
  {Schachenmayer}}, \bibinfo {author} {\bibfnamefont {C.}~\bibnamefont
  {Genes}}, \bibinfo {author} {\bibfnamefont {E.}~\bibnamefont {Tignone}},\
  and\ \bibinfo {author} {\bibfnamefont {G.}~\bibnamefont {Pupillo}},\
  }\bibfield  {title} {\bibinfo {title} {Cavity-enhanced transport of
  excitons},\ }\href {https://doi.org/10.1103/PhysRevLett.114.196403}
  {\bibfield  {journal} {\bibinfo  {journal} {Phys. Rev. Lett.}\ }\textbf
  {\bibinfo {volume} {114}},\ \bibinfo {pages} {196403} (\bibinfo {year}
  {2015})}\BibitemShut {NoStop}%
\bibitem [{\citenamefont {Melnikau}\ \emph {et~al.}(2016)\citenamefont
  {Melnikau}, \citenamefont {Esteban}, \citenamefont {Savateeva}, \citenamefont
  {S{\'a}nchez-Iglesias}, \citenamefont {Grzelczak}, \citenamefont {Schmidt},
  \citenamefont {Liz-Marz{\'a}n}, \citenamefont {Aizpurua},\ and\ \citenamefont
  {Rakovich}}]{JPhysChemLett.7.354(2016)}%
  \BibitemOpen
  \bibfield  {author} {\bibinfo {author} {\bibfnamefont {D.}~\bibnamefont
  {Melnikau}}, \bibinfo {author} {\bibfnamefont {R.}~\bibnamefont {Esteban}},
  \bibinfo {author} {\bibfnamefont {D.}~\bibnamefont {Savateeva}}, \bibinfo
  {author} {\bibfnamefont {A.}~\bibnamefont {S{\'a}nchez-Iglesias}}, \bibinfo
  {author} {\bibfnamefont {M.}~\bibnamefont {Grzelczak}}, \bibinfo {author}
  {\bibfnamefont {M.~K.}\ \bibnamefont {Schmidt}}, \bibinfo {author}
  {\bibfnamefont {L.~M.}\ \bibnamefont {Liz-Marz{\'a}n}}, \bibinfo {author}
  {\bibfnamefont {J.}~\bibnamefont {Aizpurua}},\ and\ \bibinfo {author}
  {\bibfnamefont {Y.~P.}\ \bibnamefont {Rakovich}},\ }\bibfield  {title}
  {\bibinfo {title} {Rabi splitting in photoluminescence spectra of hybrid
  systems of gold nanorods and j-aggregates},\ }\href
  {https://doi.org/10.1021/acs.jpclett.5b02512} {\bibfield  {journal} {\bibinfo
   {journal} {J. Phys. Chem. Lett.}\ }\textbf {\bibinfo {volume} {7}},\
  \bibinfo {pages} {354} (\bibinfo {year} {2016})}\BibitemShut {NoStop}%
\bibitem [{\citenamefont {Wu}\ \emph {et~al.}(2021)\citenamefont {Wu},
  \citenamefont {Guo}, \citenamefont {Huang}, \citenamefont {Liang},
  \citenamefont {Jin}, \citenamefont {Li}, \citenamefont {Deng}, \citenamefont
  {Jiao}, \citenamefont {Liu}, \citenamefont {Zhang}, \citenamefont {Zhang},\
  and\ \citenamefont {Yu}}]{ACSNano2.2292(2021)}%
  \BibitemOpen
  \bibfield  {author} {\bibinfo {author} {\bibfnamefont {F.}~\bibnamefont
  {Wu}}, \bibinfo {author} {\bibfnamefont {J.}~\bibnamefont {Guo}}, \bibinfo
  {author} {\bibfnamefont {Y.}~\bibnamefont {Huang}}, \bibinfo {author}
  {\bibfnamefont {K.}~\bibnamefont {Liang}}, \bibinfo {author} {\bibfnamefont
  {L.}~\bibnamefont {Jin}}, \bibinfo {author} {\bibfnamefont {J.}~\bibnamefont
  {Li}}, \bibinfo {author} {\bibfnamefont {X.}~\bibnamefont {Deng}}, \bibinfo
  {author} {\bibfnamefont {R.}~\bibnamefont {Jiao}}, \bibinfo {author}
  {\bibfnamefont {Y.}~\bibnamefont {Liu}}, \bibinfo {author} {\bibfnamefont
  {J.}~\bibnamefont {Zhang}}, \bibinfo {author} {\bibfnamefont
  {W.}~\bibnamefont {Zhang}},\ and\ \bibinfo {author} {\bibfnamefont
  {L.}~\bibnamefont {Yu}},\ }\bibfield  {title} {\bibinfo {title} {Plexcitonic
  optical chirality: Strong exciton--plasmon coupling in chiral
  j-aggregate-metal nanoparticle complexes},\ }\href
  {https://doi.org/10.1021/acsnano.0c08274} {\bibfield  {journal} {\bibinfo
  {journal} {ACS Nano}\ }\textbf {\bibinfo {volume} {15}},\ \bibinfo {pages}
  {2292} (\bibinfo {year} {2021})}\BibitemShut {NoStop}%
\bibitem [{\citenamefont {Feist}\ \emph {et~al.}(2018)\citenamefont {Feist},
  \citenamefont {Galego},\ and\ \citenamefont {Garcia-Vidal}}]{ACS5.205(2018)}%
  \BibitemOpen
  \bibfield  {author} {\bibinfo {author} {\bibfnamefont {J.}~\bibnamefont
  {Feist}}, \bibinfo {author} {\bibfnamefont {J.}~\bibnamefont {Galego}},\ and\
  \bibinfo {author} {\bibfnamefont {F.~J.}\ \bibnamefont {Garcia-Vidal}},\
  }\bibfield  {title} {\bibinfo {title} {Polaritonic chemistry with organic
  molecules},\ }\href {https://doi.org/10.1021/acsphotonics.7b00680} {\bibfield
   {journal} {\bibinfo  {journal} {ACS Photonics}\ }\textbf {\bibinfo {volume}
  {5}},\ \bibinfo {pages} {205} (\bibinfo {year} {2018})}\BibitemShut {NoStop}%
\bibitem [{\citenamefont {Flick}\ \emph {et~al.}(2017)\citenamefont {Flick},
  \citenamefont {Ruggenthaler}, \citenamefont {Appel},\ and\ \citenamefont
  {Rubio}}]{PNAS114.3026(2017)}%
  \BibitemOpen
  \bibfield  {author} {\bibinfo {author} {\bibfnamefont {J.}~\bibnamefont
  {Flick}}, \bibinfo {author} {\bibfnamefont {M.}~\bibnamefont {Ruggenthaler}},
  \bibinfo {author} {\bibfnamefont {H.}~\bibnamefont {Appel}},\ and\ \bibinfo
  {author} {\bibfnamefont {A.}~\bibnamefont {Rubio}},\ }\bibfield  {title}
  {\bibinfo {title} {Atoms and molecules in cavities, from weak to strong
  coupling in quantum-electrodynamics (qed) chemistry},\ }\href
  {https://doi.org/10.1073/pnas.1615509114} {\bibfield  {journal} {\bibinfo
  {journal} {P. Nat. Acad. Sci.}\ }\textbf {\bibinfo {volume} {114}},\ \bibinfo
  {pages} {3026} (\bibinfo {year} {2017})}\BibitemShut {NoStop}%
\bibitem [{\citenamefont {Herrera}\ and\ \citenamefont
  {Spano}(2016)}]{PhysRevLett116.238301(2016)}%
  \BibitemOpen
  \bibfield  {author} {\bibinfo {author} {\bibfnamefont {F.}~\bibnamefont
  {Herrera}}\ and\ \bibinfo {author} {\bibfnamefont {F.~C.}\ \bibnamefont
  {Spano}},\ }\bibfield  {title} {\bibinfo {title} {Cavity-controlled chemistry
  in molecular ensembles},\ }\href
  {https://doi.org/10.1103/PhysRevLett.116.238301} {\bibfield  {journal}
  {\bibinfo  {journal} {Phys. Rev. Lett.}\ }\textbf {\bibinfo {volume} {116}},\
  \bibinfo {pages} {238301} (\bibinfo {year} {2016})}\BibitemShut {NoStop}%
\bibitem [{\citenamefont {Vergauwe}\ \emph {et~al.}(2019)\citenamefont
  {Vergauwe}, \citenamefont {Thomas}, \citenamefont {Nagarajan}, \citenamefont
  {Shalabney}, \citenamefont {George}, \citenamefont {Chervy}, \citenamefont
  {Seidel}, \citenamefont {Devaux}, \citenamefont {Torbeev},\ and\
  \citenamefont {Ebbesen}}]{AngChem58.15324(2019)}%
  \BibitemOpen
  \bibfield  {author} {\bibinfo {author} {\bibfnamefont {R.~M.~A.}\
  \bibnamefont {Vergauwe}}, \bibinfo {author} {\bibfnamefont {A.}~\bibnamefont
  {Thomas}}, \bibinfo {author} {\bibfnamefont {K.}~\bibnamefont {Nagarajan}},
  \bibinfo {author} {\bibfnamefont {A.}~\bibnamefont {Shalabney}}, \bibinfo
  {author} {\bibfnamefont {J.}~\bibnamefont {George}}, \bibinfo {author}
  {\bibfnamefont {T.}~\bibnamefont {Chervy}}, \bibinfo {author} {\bibfnamefont
  {M.}~\bibnamefont {Seidel}}, \bibinfo {author} {\bibfnamefont
  {E.}~\bibnamefont {Devaux}}, \bibinfo {author} {\bibfnamefont
  {V.}~\bibnamefont {Torbeev}},\ and\ \bibinfo {author} {\bibfnamefont {T.~W.}\
  \bibnamefont {Ebbesen}},\ }\bibfield  {title} {\bibinfo {title} {Modification
  of enzyme activity by vibrational strong coupling of water},\ }\href
  {https://doi.org/anie.201908876} {\bibfield  {journal} {\bibinfo  {journal}
  {Angew. Chem., Int. Ed.}\ }\textbf {\bibinfo {volume} {58}},\ \bibinfo
  {pages} {15324} (\bibinfo {year} {2019})}\BibitemShut {NoStop}%
\bibitem [{\citenamefont {Thomas}\ \emph {et~al.}(2019)\citenamefont {Thomas},
  \citenamefont {Lethuillier-Karl}, \citenamefont {Nagarajan}, \citenamefont
  {Vergauwe}, \citenamefont {George}, \citenamefont {Chervy}, \citenamefont
  {Shalabney}, \citenamefont {Devaux}, \citenamefont {Genet}, \citenamefont
  {Moran},\ and\ \citenamefont {Ebbesen}}]{Science363.615(2019)}%
  \BibitemOpen
  \bibfield  {author} {\bibinfo {author} {\bibfnamefont {A.}~\bibnamefont
  {Thomas}}, \bibinfo {author} {\bibfnamefont {L.}~\bibnamefont
  {Lethuillier-Karl}}, \bibinfo {author} {\bibfnamefont {K.}~\bibnamefont
  {Nagarajan}}, \bibinfo {author} {\bibfnamefont {R.~M.~A.}\ \bibnamefont
  {Vergauwe}}, \bibinfo {author} {\bibfnamefont {J.}~\bibnamefont {George}},
  \bibinfo {author} {\bibfnamefont {T.}~\bibnamefont {Chervy}}, \bibinfo
  {author} {\bibfnamefont {A.}~\bibnamefont {Shalabney}}, \bibinfo {author}
  {\bibfnamefont {E.}~\bibnamefont {Devaux}}, \bibinfo {author} {\bibfnamefont
  {C.}~\bibnamefont {Genet}}, \bibinfo {author} {\bibfnamefont
  {J.}~\bibnamefont {Moran}},\ and\ \bibinfo {author} {\bibfnamefont {T.~W.}\
  \bibnamefont {Ebbesen}},\ }\bibfield  {title} {\bibinfo {title} {Tilting a
  ground-state reactivity landscape by vibrational strong coupling},\ }\href
  {https://doi.org/10.1126/science.aau7742} {\bibfield  {journal} {\bibinfo
  {journal} {Science}\ }\textbf {\bibinfo {volume} {363}},\ \bibinfo {pages}
  {615} (\bibinfo {year} {2019})}\BibitemShut {NoStop}%
\bibitem [{\citenamefont {Galego}\ \emph {et~al.}(2015)\citenamefont {Galego},
  \citenamefont {Garcia-Vidal},\ and\ \citenamefont
  {Feist}}]{PhysRevX.5.041022(2015)}%
  \BibitemOpen
  \bibfield  {author} {\bibinfo {author} {\bibfnamefont {J.}~\bibnamefont
  {Galego}}, \bibinfo {author} {\bibfnamefont {F.~J.}\ \bibnamefont
  {Garcia-Vidal}},\ and\ \bibinfo {author} {\bibfnamefont {J.}~\bibnamefont
  {Feist}},\ }\bibfield  {title} {\bibinfo {title} {Cavity-induced
  modifications of molecular structure in the strong-coupling regime},\ }\href
  {https://doi.org/10.1103/PhysRevX.5.041022} {\bibfield  {journal} {\bibinfo
  {journal} {Phys. Rev. X}\ }\textbf {\bibinfo {volume} {5}},\ \bibinfo {pages}
  {041022} (\bibinfo {year} {2015})}\BibitemShut {NoStop}%
\bibitem [{\citenamefont {Dutra}(2004)}]{QEDbook}%
  \BibitemOpen
  \bibfield  {author} {\bibinfo {author} {\bibfnamefont {S.~M.}\ \bibnamefont
  {Dutra}},\ }\href {https://doi.org/https://doi.org/10.1002/0471713465} {\emph
  {\bibinfo {title} {Cavity Quantum Electrodynamics}}}\ (\bibinfo  {publisher}
  {John Wiley \& Sons, Ltd},\ \bibinfo {year} {2004})\BibitemShut {NoStop}%
\bibitem [{\citenamefont {T\"orm\"a}\ and\ \citenamefont
  {Barnes}(2014)}]{RepProgPhys78.013901(2014)}%
  \BibitemOpen
  \bibfield  {author} {\bibinfo {author} {\bibfnamefont {P.}~\bibnamefont
  {T\"orm\"a}}\ and\ \bibinfo {author} {\bibfnamefont {W.~L.}\ \bibnamefont
  {Barnes}},\ }\bibfield  {title} {\bibinfo {title} {Strong coupling between
  surface plasmon polaritons and emitters: a review},\ }\href
  {https://doi.org/10.1088/0034-4885/78/1/013901} {\bibfield  {journal}
  {\bibinfo  {journal} {Rep. Prog. Phys.}\ }\textbf {\bibinfo {volume} {78}},\
  \bibinfo {pages} {013901} (\bibinfo {year} {2014})}\BibitemShut {NoStop}%
\bibitem [{\citenamefont {Ebbesen}(2016)}]{ACS49.2403(2016)}%
  \BibitemOpen
  \bibfield  {author} {\bibinfo {author} {\bibfnamefont {T.~W.}\ \bibnamefont
  {Ebbesen}},\ }\bibfield  {title} {\bibinfo {title} {Hybrid light--matter
  states in a molecular and material science perspective},\ }\href
  {https://doi.org/10.1021/acs.accounts.6b00295} {\bibfield  {journal}
  {\bibinfo  {journal} {Accounts Chem. Res.}\ }\textbf {\bibinfo {volume}
  {49}},\ \bibinfo {pages} {2403} (\bibinfo {year} {2016})}\BibitemShut
  {NoStop}%
\bibitem [{\citenamefont {Xiang}\ and\ \citenamefont
  {Xiong}(2021)}]{JChemPhys155.050901(2021)}%
  \BibitemOpen
  \bibfield  {author} {\bibinfo {author} {\bibfnamefont {B.}~\bibnamefont
  {Xiang}}\ and\ \bibinfo {author} {\bibfnamefont {W.}~\bibnamefont {Xiong}},\
  }\bibfield  {title} {\bibinfo {title} {Molecular vibrational polariton: Its
  dynamics and potentials in novel chemistry and quantum technology},\ }\href
  {https://doi.org/10.1063/5.0054896} {\bibfield  {journal} {\bibinfo
  {journal} {J. Chem. Phys.}\ }\textbf {\bibinfo {volume} {155}},\ \bibinfo
  {pages} {050901} (\bibinfo {year} {2021})}\BibitemShut {NoStop}%
\bibitem [{\citenamefont {Xiang}\ \emph {et~al.}(2018)\citenamefont {Xiang},
  \citenamefont {Ribeiro}, \citenamefont {Dunkelberger}, \citenamefont {Wang},
  \citenamefont {Li}, \citenamefont {Simpkins}, \citenamefont {Owrutsky},
  \citenamefont {Yuen-Zhou},\ and\ \citenamefont {Xiong}}]{PNAS115.4845(2018)}%
  \BibitemOpen
  \bibfield  {author} {\bibinfo {author} {\bibfnamefont {B.}~\bibnamefont
  {Xiang}}, \bibinfo {author} {\bibfnamefont {R.~F.}\ \bibnamefont {Ribeiro}},
  \bibinfo {author} {\bibfnamefont {A.~D.}\ \bibnamefont {Dunkelberger}},
  \bibinfo {author} {\bibfnamefont {J.}~\bibnamefont {Wang}}, \bibinfo {author}
  {\bibfnamefont {Y.}~\bibnamefont {Li}}, \bibinfo {author} {\bibfnamefont
  {B.~S.}\ \bibnamefont {Simpkins}}, \bibinfo {author} {\bibfnamefont {J.~C.}\
  \bibnamefont {Owrutsky}}, \bibinfo {author} {\bibfnamefont {J.}~\bibnamefont
  {Yuen-Zhou}},\ and\ \bibinfo {author} {\bibfnamefont {W.}~\bibnamefont
  {Xiong}},\ }\bibfield  {title} {\bibinfo {title} {Two-dimensional infrared
  spectroscopy of vibrational polaritons},\ }\href
  {https://doi.org/10.1073/pnas.1722063115} {\bibfield  {journal} {\bibinfo
  {journal} {P. Natl. Acad. Sci.}\ }\textbf {\bibinfo {volume} {115}},\
  \bibinfo {pages} {4845} (\bibinfo {year} {2018})}\BibitemShut {NoStop}%
\bibitem [{\citenamefont {Benisty}\ \emph {et~al.}(2022)\citenamefont
  {Benisty}, \citenamefont {Greffet},\ and\ \citenamefont
  {Lalanne}}]{IntNanophotonics}%
  \BibitemOpen
  \bibfield  {author} {\bibinfo {author} {\bibfnamefont {H.}~\bibnamefont
  {Benisty}}, \bibinfo {author} {\bibfnamefont {J.-J.}\ \bibnamefont
  {Greffet}},\ and\ \bibinfo {author} {\bibfnamefont {P.}~\bibnamefont
  {Lalanne}},\ }\href@noop {} {\emph {\bibinfo {title} {Introduction to
  Nanophotonics}}}\ (\bibinfo  {publisher} {Oxford Graduate Texts},\ \bibinfo
  {year} {2022})\BibitemShut {NoStop}%
\bibitem [{\citenamefont {Barra-Burillo}\ \emph {et~al.}(2021)\citenamefont
  {Barra-Burillo}, \citenamefont {Muniain}, \citenamefont {Catalano},
  \citenamefont {Autore}, \citenamefont {Casanova}, \citenamefont {Hueso},
  \citenamefont {Aizpurua}, \citenamefont {Esteban},\ and\ \citenamefont
  {Hillenbrand}}]{NatComm12.6206(2021)}%
  \BibitemOpen
  \bibfield  {author} {\bibinfo {author} {\bibfnamefont {M.}~\bibnamefont
  {Barra-Burillo}}, \bibinfo {author} {\bibfnamefont {U.}~\bibnamefont
  {Muniain}}, \bibinfo {author} {\bibfnamefont {S.}~\bibnamefont {Catalano}},
  \bibinfo {author} {\bibfnamefont {M.}~\bibnamefont {Autore}}, \bibinfo
  {author} {\bibfnamefont {F.}~\bibnamefont {Casanova}}, \bibinfo {author}
  {\bibfnamefont {L.~E.}\ \bibnamefont {Hueso}}, \bibinfo {author}
  {\bibfnamefont {J.}~\bibnamefont {Aizpurua}}, \bibinfo {author}
  {\bibfnamefont {R.}~\bibnamefont {Esteban}},\ and\ \bibinfo {author}
  {\bibfnamefont {R.}~\bibnamefont {Hillenbrand}},\ }\bibfield  {title}
  {\bibinfo {title} {Microcavity phonon polaritons from the weak to the
  ultrastrong phonon--photon coupling regime},\ }\href
  {https://doi.org/10.1038/s41467-021-26060-x} {\bibfield  {journal} {\bibinfo
  {journal} {Nat. Comm.}\ }\textbf {\bibinfo {volume} {12}},\ \bibinfo {pages}
  {6206} (\bibinfo {year} {2021})}\BibitemShut {NoStop}%
\bibitem [{\citenamefont {Schilder}\ \emph {et~al.}(2016)\citenamefont
  {Schilder}, \citenamefont {Sauvan}, \citenamefont {Hugonin}, \citenamefont
  {Jennewein}, \citenamefont {Sortais}, \citenamefont {Browaeys},\ and\
  \citenamefont {Greffet}}]{PhysRevA93.063835(2016)}%
  \BibitemOpen
  \bibfield  {author} {\bibinfo {author} {\bibfnamefont {N.~J.}\ \bibnamefont
  {Schilder}}, \bibinfo {author} {\bibfnamefont {C.}~\bibnamefont {Sauvan}},
  \bibinfo {author} {\bibfnamefont {J.-P.}\ \bibnamefont {Hugonin}}, \bibinfo
  {author} {\bibfnamefont {S.}~\bibnamefont {Jennewein}}, \bibinfo {author}
  {\bibfnamefont {Y.~R.~P.}\ \bibnamefont {Sortais}}, \bibinfo {author}
  {\bibfnamefont {A.}~\bibnamefont {Browaeys}},\ and\ \bibinfo {author}
  {\bibfnamefont {J.-J.}\ \bibnamefont {Greffet}},\ }\bibfield  {title}
  {\bibinfo {title} {Polaritonic modes in a dense cloud of cold atoms},\ }\href
  {https://doi.org/10.1103/PhysRevA.93.063835} {\bibfield  {journal} {\bibinfo
  {journal} {Phys. Rev. A}\ }\textbf {\bibinfo {volume} {93}},\ \bibinfo
  {pages} {063835} (\bibinfo {year} {2016})}\BibitemShut {NoStop}%
\bibitem [{\citenamefont {Jennewein}\ \emph {et~al.}(2016)\citenamefont
  {Jennewein}, \citenamefont {Besbes}, \citenamefont {Schilder}, \citenamefont
  {Jenkins}, \citenamefont {Sauvan}, \citenamefont {Ruostekoski}, \citenamefont
  {Greffet}, \citenamefont {Sortais},\ and\ \citenamefont
  {Browaeys}}]{PhysRevLett116.233601(2016)}%
  \BibitemOpen
  \bibfield  {author} {\bibinfo {author} {\bibfnamefont {S.}~\bibnamefont
  {Jennewein}}, \bibinfo {author} {\bibfnamefont {M.}~\bibnamefont {Besbes}},
  \bibinfo {author} {\bibfnamefont {N.~J.}\ \bibnamefont {Schilder}}, \bibinfo
  {author} {\bibfnamefont {S.~D.}\ \bibnamefont {Jenkins}}, \bibinfo {author}
  {\bibfnamefont {C.}~\bibnamefont {Sauvan}}, \bibinfo {author} {\bibfnamefont
  {J.}~\bibnamefont {Ruostekoski}}, \bibinfo {author} {\bibfnamefont {J.-J.}\
  \bibnamefont {Greffet}}, \bibinfo {author} {\bibfnamefont {Y.~R.~P.}\
  \bibnamefont {Sortais}},\ and\ \bibinfo {author} {\bibfnamefont
  {A.}~\bibnamefont {Browaeys}},\ }\bibfield  {title} {\bibinfo {title}
  {Coherent scattering of near-resonant light by a dense microscopic cold
  atomic cloud},\ }\href {https://doi.org/10.1103/PhysRevLett.116.233601}
  {\bibfield  {journal} {\bibinfo  {journal} {Phys. Rev. Lett.}\ }\textbf
  {\bibinfo {volume} {116}},\ \bibinfo {pages} {233601} (\bibinfo {year}
  {2016})}\BibitemShut {NoStop}%
\bibitem [{\citenamefont {Basov}\ \emph {et~al.}(2016)\citenamefont {Basov},
  \citenamefont {Fogler},\ and\ \citenamefont
  {de~Abajo}}]{Science6309.1992(2016)}%
  \BibitemOpen
  \bibfield  {author} {\bibinfo {author} {\bibfnamefont {D.~N.}\ \bibnamefont
  {Basov}}, \bibinfo {author} {\bibfnamefont {M.~M.}\ \bibnamefont {Fogler}},\
  and\ \bibinfo {author} {\bibfnamefont {F.~J.~G.}\ \bibnamefont {de~Abajo}},\
  }\bibfield  {title} {\bibinfo {title} {Polaritons in van der waals
  materials},\ }\href {https://doi.org/10.1126/science.aag1992} {\bibfield
  {journal} {\bibinfo  {journal} {Science}\ }\textbf {\bibinfo {volume}
  {354}},\ \bibinfo {pages} {aag1992} (\bibinfo {year} {2016})}\BibitemShut
  {NoStop}%
\bibitem [{\citenamefont {Autore}\ \emph {et~al.}(2018)\citenamefont {Autore},
  \citenamefont {Li}, \citenamefont {Dolado}, \citenamefont {Alfaro-Mozaz},
  \citenamefont {Esteban}, \citenamefont {Atxabal}, \citenamefont {Casanova},
  \citenamefont {Hueso}, \citenamefont {Alonso-Gonz{\'a}lez}, \citenamefont
  {Aizpurua}, \citenamefont {Nikitin}, \citenamefont {V{\'e}lez},\ and\
  \citenamefont {Hillenbrand}}]{LightSciAppl7.17172(2018)}%
  \BibitemOpen
  \bibfield  {author} {\bibinfo {author} {\bibfnamefont {M.}~\bibnamefont
  {Autore}}, \bibinfo {author} {\bibfnamefont {P.}~\bibnamefont {Li}}, \bibinfo
  {author} {\bibfnamefont {I.}~\bibnamefont {Dolado}}, \bibinfo {author}
  {\bibfnamefont {F.~J.}\ \bibnamefont {Alfaro-Mozaz}}, \bibinfo {author}
  {\bibfnamefont {R.}~\bibnamefont {Esteban}}, \bibinfo {author} {\bibfnamefont
  {A.}~\bibnamefont {Atxabal}}, \bibinfo {author} {\bibfnamefont
  {F.}~\bibnamefont {Casanova}}, \bibinfo {author} {\bibfnamefont {L.~E.}\
  \bibnamefont {Hueso}}, \bibinfo {author} {\bibfnamefont {P.}~\bibnamefont
  {Alonso-Gonz{\'a}lez}}, \bibinfo {author} {\bibfnamefont {J.}~\bibnamefont
  {Aizpurua}}, \bibinfo {author} {\bibfnamefont {A.~Y.}\ \bibnamefont
  {Nikitin}}, \bibinfo {author} {\bibfnamefont {S.}~\bibnamefont {V{\'e}lez}},\
  and\ \bibinfo {author} {\bibfnamefont {R.}~\bibnamefont {Hillenbrand}},\
  }\bibfield  {title} {\bibinfo {title} {Boron nitride nanoresonators for
  phonon-enhanced molecular vibrational spectroscopy at the strong coupling
  limit},\ }\href {https://doi.org/10.1038/lsa.2017.172} {\bibfield  {journal}
  {\bibinfo  {journal} {Light: Science {\&} Applications}\ }\textbf {\bibinfo
  {volume} {7}},\ \bibinfo {pages} {17172} (\bibinfo {year}
  {2018})}\BibitemShut {NoStop}%
\bibitem [{\citenamefont {Ma}\ \emph {et~al.}(2018)\citenamefont {Ma},
  \citenamefont {Alonso-Gonz{\'a}lez}, \citenamefont {Li}, \citenamefont
  {Nikitin}, \citenamefont {Yuan}, \citenamefont {Mart{\'i}n-S{\'a}nchez},
  \citenamefont {Taboada-Guti{\'e}rrez}, \citenamefont {Amenabar},
  \citenamefont {Li}, \citenamefont {V{\'e}lez}, \citenamefont {Tollan},
  \citenamefont {Dai}, \citenamefont {Zhang}, \citenamefont {Sriram},
  \citenamefont {Kalantar-Zadeh}, \citenamefont {Lee}, \citenamefont
  {Hillenbrand},\ and\ \citenamefont {Bao}}]{Nature562.557(2018)}%
  \BibitemOpen
  \bibfield  {author} {\bibinfo {author} {\bibfnamefont {W.}~\bibnamefont
  {Ma}}, \bibinfo {author} {\bibfnamefont {P.}~\bibnamefont
  {Alonso-Gonz{\'a}lez}}, \bibinfo {author} {\bibfnamefont {S.}~\bibnamefont
  {Li}}, \bibinfo {author} {\bibfnamefont {A.~Y.}\ \bibnamefont {Nikitin}},
  \bibinfo {author} {\bibfnamefont {J.}~\bibnamefont {Yuan}}, \bibinfo {author}
  {\bibfnamefont {J.}~\bibnamefont {Mart{\'i}n-S{\'a}nchez}}, \bibinfo {author}
  {\bibfnamefont {J.}~\bibnamefont {Taboada-Guti{\'e}rrez}}, \bibinfo {author}
  {\bibfnamefont {I.}~\bibnamefont {Amenabar}}, \bibinfo {author}
  {\bibfnamefont {P.}~\bibnamefont {Li}}, \bibinfo {author} {\bibfnamefont
  {S.}~\bibnamefont {V{\'e}lez}}, \bibinfo {author} {\bibfnamefont
  {C.}~\bibnamefont {Tollan}}, \bibinfo {author} {\bibfnamefont
  {Z.}~\bibnamefont {Dai}}, \bibinfo {author} {\bibfnamefont {Y.}~\bibnamefont
  {Zhang}}, \bibinfo {author} {\bibfnamefont {S.}~\bibnamefont {Sriram}},
  \bibinfo {author} {\bibfnamefont {K.}~\bibnamefont {Kalantar-Zadeh}},
  \bibinfo {author} {\bibfnamefont {S.-T.}\ \bibnamefont {Lee}}, \bibinfo
  {author} {\bibfnamefont {R.}~\bibnamefont {Hillenbrand}},\ and\ \bibinfo
  {author} {\bibfnamefont {Q.}~\bibnamefont {Bao}},\ }\bibfield  {title}
  {\bibinfo {title} {In-plane anisotropic and ultra-low-loss polaritons in a
  natural van der waals crystal},\ }\href
  {https://doi.org/10.1038/s41586-018-0618-9} {\bibfield  {journal} {\bibinfo
  {journal} {Nature}\ }\textbf {\bibinfo {volume} {562}},\ \bibinfo {pages}
  {557} (\bibinfo {year} {2018})}\BibitemShut {NoStop}%
\bibitem [{\citenamefont {Iranzo}\ \emph {et~al.}(2018)\citenamefont {Iranzo},
  \citenamefont {Nanot}, \citenamefont {Dias}, \citenamefont {Epstein},
  \citenamefont {Peng}, \citenamefont {Efetov}, \citenamefont {Lundeberg},
  \citenamefont {Parret}, \citenamefont {Osmond}, \citenamefont {Hong},
  \citenamefont {Kong}, \citenamefont {Englund}, \citenamefont {Peres},\ and\
  \citenamefont {Koppens}}]{Science6386.291(2018)}%
  \BibitemOpen
  \bibfield  {author} {\bibinfo {author} {\bibfnamefont {D.~A.}\ \bibnamefont
  {Iranzo}}, \bibinfo {author} {\bibfnamefont {S.}~\bibnamefont {Nanot}},
  \bibinfo {author} {\bibfnamefont {E.~J.~C.}\ \bibnamefont {Dias}}, \bibinfo
  {author} {\bibfnamefont {I.}~\bibnamefont {Epstein}}, \bibinfo {author}
  {\bibfnamefont {C.}~\bibnamefont {Peng}}, \bibinfo {author} {\bibfnamefont
  {D.~K.}\ \bibnamefont {Efetov}}, \bibinfo {author} {\bibfnamefont {M.~B.}\
  \bibnamefont {Lundeberg}}, \bibinfo {author} {\bibfnamefont {R.}~\bibnamefont
  {Parret}}, \bibinfo {author} {\bibfnamefont {J.}~\bibnamefont {Osmond}},
  \bibinfo {author} {\bibfnamefont {J.-Y.}\ \bibnamefont {Hong}}, \bibinfo
  {author} {\bibfnamefont {J.}~\bibnamefont {Kong}}, \bibinfo {author}
  {\bibfnamefont {D.~R.}\ \bibnamefont {Englund}}, \bibinfo {author}
  {\bibfnamefont {N.~M.~R.}\ \bibnamefont {Peres}},\ and\ \bibinfo {author}
  {\bibfnamefont {F.~H.~L.}\ \bibnamefont {Koppens}},\ }\bibfield  {title}
  {\bibinfo {title} {Probing the ultimate plasmon confinement limits with a van
  der waals heterostructure},\ }\href {https://doi.org/10.1126/science.aar8438}
  {\bibfield  {journal} {\bibinfo  {journal} {Science}\ }\textbf {\bibinfo
  {volume} {360}},\ \bibinfo {pages} {291} (\bibinfo {year}
  {2018})}\BibitemShut {NoStop}%
\bibitem [{\citenamefont {Singh}\ \emph {et~al.}(2020)\citenamefont {Singh},
  \citenamefont {Kaur},\ and\ \citenamefont {Comini}}]{RSC8.3938(2020)}%
  \BibitemOpen
  \bibfield  {author} {\bibinfo {author} {\bibfnamefont {M.}~\bibnamefont
  {Singh}}, \bibinfo {author} {\bibfnamefont {N.}~\bibnamefont {Kaur}},\ and\
  \bibinfo {author} {\bibfnamefont {E.}~\bibnamefont {Comini}},\ }\bibfield
  {title} {\bibinfo {title} {The role of self-assembled monolayers in
  electronic devices},\ }\href {https://doi.org/10.1039/D0TC00388C} {\bibfield
  {journal} {\bibinfo  {journal} {J. Mater. Chem. C}\ }\textbf {\bibinfo
  {volume} {8}},\ \bibinfo {pages} {3938} (\bibinfo {year} {2020})}\BibitemShut
  {NoStop}%
\bibitem [{\citenamefont {Zhang}\ \emph {et~al.}(2013)\citenamefont {Zhang},
  \citenamefont {Zhang}, \citenamefont {Dong}, \citenamefont {Jiang},
  \citenamefont {Zhang}, \citenamefont {Chen}, \citenamefont {Zhang},
  \citenamefont {Liao}, \citenamefont {Aizpurua}, \citenamefont {Luo},
  \citenamefont {Yang},\ and\ \citenamefont {Hou}}]{Nature498.82(2013)}%
  \BibitemOpen
  \bibfield  {author} {\bibinfo {author} {\bibfnamefont {R.}~\bibnamefont
  {Zhang}}, \bibinfo {author} {\bibfnamefont {Y.}~\bibnamefont {Zhang}},
  \bibinfo {author} {\bibfnamefont {Z.~C.}\ \bibnamefont {Dong}}, \bibinfo
  {author} {\bibfnamefont {S.}~\bibnamefont {Jiang}}, \bibinfo {author}
  {\bibfnamefont {C.}~\bibnamefont {Zhang}}, \bibinfo {author} {\bibfnamefont
  {L.~G.}\ \bibnamefont {Chen}}, \bibinfo {author} {\bibfnamefont
  {L.}~\bibnamefont {Zhang}}, \bibinfo {author} {\bibfnamefont
  {Y.}~\bibnamefont {Liao}}, \bibinfo {author} {\bibfnamefont {J.}~\bibnamefont
  {Aizpurua}}, \bibinfo {author} {\bibfnamefont {Y.}~\bibnamefont {Luo}},
  \bibinfo {author} {\bibfnamefont {J.~L.}\ \bibnamefont {Yang}},\ and\
  \bibinfo {author} {\bibfnamefont {J.~G.}\ \bibnamefont {Hou}},\ }\bibfield
  {title} {\bibinfo {title} {Chemical mapping of a single molecule by
  plasmon-enhanced raman scattering},\ }\href
  {https://doi.org/10.1038/nature12151} {\bibfield  {journal} {\bibinfo
  {journal} {Nature}\ }\textbf {\bibinfo {volume} {498}},\ \bibinfo {pages}
  {82} (\bibinfo {year} {2013})}\BibitemShut {NoStop}%
\bibitem [{\citenamefont {Barbry}\ \emph {et~al.}(2015)\citenamefont {Barbry},
  \citenamefont {Koval}, \citenamefont {Marchesin}, \citenamefont {Esteban},
  \citenamefont {Borisov}, \citenamefont {Aizpurua},\ and\ \citenamefont
  {S{\'a}nchez-Portal}}]{NanoLett5.3410(2015)}%
  \BibitemOpen
  \bibfield  {author} {\bibinfo {author} {\bibfnamefont {M.}~\bibnamefont
  {Barbry}}, \bibinfo {author} {\bibfnamefont {P.}~\bibnamefont {Koval}},
  \bibinfo {author} {\bibfnamefont {F.}~\bibnamefont {Marchesin}}, \bibinfo
  {author} {\bibfnamefont {R.}~\bibnamefont {Esteban}}, \bibinfo {author}
  {\bibfnamefont {A.~G.}\ \bibnamefont {Borisov}}, \bibinfo {author}
  {\bibfnamefont {J.}~\bibnamefont {Aizpurua}},\ and\ \bibinfo {author}
  {\bibfnamefont {D.}~\bibnamefont {S{\'a}nchez-Portal}},\ }\bibfield  {title}
  {\bibinfo {title} {Atomistic near-field nanoplasmonics: Reaching atomic-scale
  resolution in nanooptics},\ }\href
  {https://doi.org/10.1021/acs.nanolett.5b00759} {\bibfield  {journal}
  {\bibinfo  {journal} {Nano Lett.}\ }\textbf {\bibinfo {volume} {15}},\
  \bibinfo {pages} {3410} (\bibinfo {year} {2015})}\BibitemShut {NoStop}%
\bibitem [{\citenamefont {Benz}\ \emph {et~al.}(2016)\citenamefont {Benz},
  \citenamefont {Schmidt}, \citenamefont {Dreismann}, \citenamefont
  {Chikkaraddy}, \citenamefont {Zhang}, \citenamefont {Demetriadou},
  \citenamefont {Carnegie}, \citenamefont {Ohadi}, \citenamefont {de~Nijs},
  \citenamefont {Esteban}, \citenamefont {Aizpurua},\ and\ \citenamefont
  {Baumberg}}]{Science6313.726(2016)}%
  \BibitemOpen
  \bibfield  {author} {\bibinfo {author} {\bibfnamefont {F.}~\bibnamefont
  {Benz}}, \bibinfo {author} {\bibfnamefont {M.~K.}\ \bibnamefont {Schmidt}},
  \bibinfo {author} {\bibfnamefont {A.}~\bibnamefont {Dreismann}}, \bibinfo
  {author} {\bibfnamefont {R.}~\bibnamefont {Chikkaraddy}}, \bibinfo {author}
  {\bibfnamefont {Y.}~\bibnamefont {Zhang}}, \bibinfo {author} {\bibfnamefont
  {A.}~\bibnamefont {Demetriadou}}, \bibinfo {author} {\bibfnamefont
  {C.}~\bibnamefont {Carnegie}}, \bibinfo {author} {\bibfnamefont
  {H.}~\bibnamefont {Ohadi}}, \bibinfo {author} {\bibfnamefont
  {B.}~\bibnamefont {de~Nijs}}, \bibinfo {author} {\bibfnamefont
  {R.}~\bibnamefont {Esteban}}, \bibinfo {author} {\bibfnamefont
  {J.}~\bibnamefont {Aizpurua}},\ and\ \bibinfo {author} {\bibfnamefont
  {J.~J.}\ \bibnamefont {Baumberg}},\ }\bibfield  {title} {\bibinfo {title}
  {Single-molecule optomechanics in picocavities},\ }\href
  {https://doi.org/10.1126/science.aah5243} {\bibfield  {journal} {\bibinfo
  {journal} {Science}\ }\textbf {\bibinfo {volume} {354}},\ \bibinfo {pages}
  {726} (\bibinfo {year} {2016})}\BibitemShut {NoStop}%
\bibitem [{\citenamefont {Doppagne}\ \emph {et~al.}(2018)\citenamefont
  {Doppagne}, \citenamefont {Chong}, \citenamefont {Bulou}, \citenamefont
  {Boeglin}, \citenamefont {Scheurer},\ and\ \citenamefont
  {Schull}}]{Science6399.251(2018)}%
  \BibitemOpen
  \bibfield  {author} {\bibinfo {author} {\bibfnamefont {B.}~\bibnamefont
  {Doppagne}}, \bibinfo {author} {\bibfnamefont {M.~C.}\ \bibnamefont {Chong}},
  \bibinfo {author} {\bibfnamefont {H.}~\bibnamefont {Bulou}}, \bibinfo
  {author} {\bibfnamefont {A.}~\bibnamefont {Boeglin}}, \bibinfo {author}
  {\bibfnamefont {F.}~\bibnamefont {Scheurer}},\ and\ \bibinfo {author}
  {\bibfnamefont {G.}~\bibnamefont {Schull}},\ }\bibfield  {title} {\bibinfo
  {title} {Electrofluorochromism at the single-molecule level},\ }\href
  {https://doi.org/10.1126/science.aat1603} {\bibfield  {journal} {\bibinfo
  {journal} {Science}\ }\textbf {\bibinfo {volume} {361}},\ \bibinfo {pages}
  {251} (\bibinfo {year} {2018})}\BibitemShut {NoStop}%
\bibitem [{\citenamefont {Jakob}\ \emph {et~al.}(2022)\citenamefont {Jakob},
  \citenamefont {Deacon}, \citenamefont {Zhang}, \citenamefont {de~Nijs},
  \citenamefont {Pavlenko}, \citenamefont {Hu}, \citenamefont {Carnegie},
  \citenamefont {Neuman}, \citenamefont {Esteban}, \citenamefont {Aizpurua},\
  and\ \citenamefont {Baumberg}}]{arxiv2204.09641}%
  \BibitemOpen
  \bibfield  {author} {\bibinfo {author} {\bibfnamefont {L.~A.}\ \bibnamefont
  {Jakob}}, \bibinfo {author} {\bibfnamefont {W.~M.}\ \bibnamefont {Deacon}},
  \bibinfo {author} {\bibfnamefont {Y.}~\bibnamefont {Zhang}}, \bibinfo
  {author} {\bibfnamefont {B.}~\bibnamefont {de~Nijs}}, \bibinfo {author}
  {\bibfnamefont {E.}~\bibnamefont {Pavlenko}}, \bibinfo {author}
  {\bibfnamefont {S.}~\bibnamefont {Hu}}, \bibinfo {author} {\bibfnamefont
  {C.}~\bibnamefont {Carnegie}}, \bibinfo {author} {\bibfnamefont
  {T.}~\bibnamefont {Neuman}}, \bibinfo {author} {\bibfnamefont
  {R.}~\bibnamefont {Esteban}}, \bibinfo {author} {\bibfnamefont
  {J.}~\bibnamefont {Aizpurua}},\ and\ \bibinfo {author} {\bibfnamefont
  {J.~J.}\ \bibnamefont {Baumberg}},\ }\href@noop {} {\bibinfo {title}
  {Softening molecular bonds through the giant optomechanical spring effect in
  plasmonic nanocavities}} (\bibinfo {year} {2022}),\ \Eprint
  {https://arxiv.org/abs/2204.09641} {arXiv:2204.09641 [physics.optics]}
  \BibitemShut {NoStop}%
\bibitem [{\citenamefont {Doppagne}\ \emph {et~al.}(2017)\citenamefont
  {Doppagne}, \citenamefont {Chong}, \citenamefont {Lorchat}, \citenamefont
  {Berciaud}, \citenamefont {Romeo}, \citenamefont {Bulou}, \citenamefont
  {Boeglin}, \citenamefont {Scheurer},\ and\ \citenamefont
  {Schull}}]{PhysRevLett.118.127401(2017)}%
  \BibitemOpen
  \bibfield  {author} {\bibinfo {author} {\bibfnamefont {B.}~\bibnamefont
  {Doppagne}}, \bibinfo {author} {\bibfnamefont {M.~C.}\ \bibnamefont {Chong}},
  \bibinfo {author} {\bibfnamefont {E.}~\bibnamefont {Lorchat}}, \bibinfo
  {author} {\bibfnamefont {S.}~\bibnamefont {Berciaud}}, \bibinfo {author}
  {\bibfnamefont {M.}~\bibnamefont {Romeo}}, \bibinfo {author} {\bibfnamefont
  {H.}~\bibnamefont {Bulou}}, \bibinfo {author} {\bibfnamefont
  {A.}~\bibnamefont {Boeglin}}, \bibinfo {author} {\bibfnamefont
  {F.}~\bibnamefont {Scheurer}},\ and\ \bibinfo {author} {\bibfnamefont
  {G.}~\bibnamefont {Schull}},\ }\bibfield  {title} {\bibinfo {title} {Vibronic
  spectroscopy with submolecular resolution from stm-induced
  electroluminescence},\ }\href
  {https://doi.org/10.1103/PhysRevLett.118.127401} {\bibfield  {journal}
  {\bibinfo  {journal} {Phys. Rev. Lett.}\ }\textbf {\bibinfo {volume} {118}},\
  \bibinfo {pages} {127401} (\bibinfo {year} {2017})}\BibitemShut {NoStop}%
\bibitem [{\citenamefont {Frisk~Kockum}\ \emph {et~al.}(2019)\citenamefont
  {Frisk~Kockum}, \citenamefont {Miranowicz}, \citenamefont {De~Liberato},
  \citenamefont {Savasta},\ and\ \citenamefont {Nori}}]{NatRevPhys1.19(2019)}%
  \BibitemOpen
  \bibfield  {author} {\bibinfo {author} {\bibfnamefont {A.}~\bibnamefont
  {Frisk~Kockum}}, \bibinfo {author} {\bibfnamefont {A.}~\bibnamefont
  {Miranowicz}}, \bibinfo {author} {\bibfnamefont {S.}~\bibnamefont
  {De~Liberato}}, \bibinfo {author} {\bibfnamefont {S.}~\bibnamefont
  {Savasta}},\ and\ \bibinfo {author} {\bibfnamefont {F.}~\bibnamefont
  {Nori}},\ }\bibfield  {title} {\bibinfo {title} {Ultrastrong coupling between
  light and matter},\ }\href {https://doi.org/10.1038/s42254-018-0006-2}
  {\bibfield  {journal} {\bibinfo  {journal} {Nat. Rev. Phys.}\ }\textbf
  {\bibinfo {volume} {1}},\ \bibinfo {pages} {19} (\bibinfo {year}
  {2019})}\BibitemShut {NoStop}%
\bibitem [{\citenamefont {Tufarelli}\ \emph {et~al.}(2015)\citenamefont
  {Tufarelli}, \citenamefont {McEnery}, \citenamefont {Maier},\ and\
  \citenamefont {Kim}}]{PhysRevA91.063840(2015)}%
  \BibitemOpen
  \bibfield  {author} {\bibinfo {author} {\bibfnamefont {T.}~\bibnamefont
  {Tufarelli}}, \bibinfo {author} {\bibfnamefont {K.~R.}\ \bibnamefont
  {McEnery}}, \bibinfo {author} {\bibfnamefont {S.~A.}\ \bibnamefont {Maier}},\
  and\ \bibinfo {author} {\bibfnamefont {M.~S.}\ \bibnamefont {Kim}},\
  }\bibfield  {title} {\bibinfo {title} {Signatures of the $a^{2}$ term in
  ultrastrongly coupled oscillators},\ }\href
  {https://doi.org/10.1103/PhysRevA.91.063840} {\bibfield  {journal} {\bibinfo
  {journal} {Phys. Rev. A}\ }\textbf {\bibinfo {volume} {91}},\ \bibinfo
  {pages} {063840} (\bibinfo {year} {2015})}\BibitemShut {NoStop}%
\bibitem [{\citenamefont {Bogoljubov}(1958)}]{Bogoliubov}%
  \BibitemOpen
  \bibfield  {author} {\bibinfo {author} {\bibfnamefont {N.~N.}\ \bibnamefont
  {Bogoljubov}},\ }\bibfield  {title} {\bibinfo {title} {On a new method in the
  theory of superconductivity},\ }\href {https://doi.org/10.1007/BF02745585}
  {\bibfield  {journal} {\bibinfo  {journal} {Il Nuovo Cimento (1955-1965)}\
  }\textbf {\bibinfo {volume} {7}},\ \bibinfo {pages} {794} (\bibinfo {year}
  {1958})}\BibitemShut {NoStop}%
\bibitem [{\citenamefont {Hopfield}(1958)}]{PhysRev112.1555(1958)}%
  \BibitemOpen
  \bibfield  {author} {\bibinfo {author} {\bibfnamefont {J.~J.}\ \bibnamefont
  {Hopfield}},\ }\bibfield  {title} {\bibinfo {title} {Theory of the
  contribution of excitons to the complex dielectric constant of crystals},\
  }\href {https://doi.org/10.1103/PhysRev.112.1555} {\bibfield  {journal}
  {\bibinfo  {journal} {Phys. Rev.}\ }\textbf {\bibinfo {volume} {112}},\
  \bibinfo {pages} {1555} (\bibinfo {year} {1958})}\BibitemShut {NoStop}%
\bibitem [{\citenamefont {Todorov}\ \emph {et~al.}(2012)\citenamefont
  {Todorov}, \citenamefont {Tosetto}, \citenamefont {Delteil}, \citenamefont
  {Vasanelli}, \citenamefont {Sirtori}, \citenamefont {Andrews},\ and\
  \citenamefont {Strasser}}]{PRB86.125314(2012)}%
  \BibitemOpen
  \bibfield  {author} {\bibinfo {author} {\bibfnamefont {Y.}~\bibnamefont
  {Todorov}}, \bibinfo {author} {\bibfnamefont {L.}~\bibnamefont {Tosetto}},
  \bibinfo {author} {\bibfnamefont {A.}~\bibnamefont {Delteil}}, \bibinfo
  {author} {\bibfnamefont {A.}~\bibnamefont {Vasanelli}}, \bibinfo {author}
  {\bibfnamefont {C.}~\bibnamefont {Sirtori}}, \bibinfo {author} {\bibfnamefont
  {A.~M.}\ \bibnamefont {Andrews}},\ and\ \bibinfo {author} {\bibfnamefont
  {G.}~\bibnamefont {Strasser}},\ }\bibfield  {title} {\bibinfo {title}
  {Polaritonic spectroscopy of intersubband transitions},\ }\href
  {https://doi.org/10.1103/PhysRevB.86.125314} {\bibfield  {journal} {\bibinfo
  {journal} {Phys. Rev. B}\ }\textbf {\bibinfo {volume} {86}},\ \bibinfo
  {pages} {125314} (\bibinfo {year} {2012})}\BibitemShut {NoStop}%
\bibitem [{\citenamefont {Hennessy}\ \emph {et~al.}(2007)\citenamefont
  {Hennessy}, \citenamefont {Badolato}, \citenamefont {Winger}, \citenamefont
  {Gerace}, \citenamefont {Atat{\"u}re}, \citenamefont {Gulde}, \citenamefont
  {F{\"a}lt}, \citenamefont {Hu},\ and\ \citenamefont
  {Imamo{\u{g}}lu}}]{Nature445.896(2007)}%
  \BibitemOpen
  \bibfield  {author} {\bibinfo {author} {\bibfnamefont {K.}~\bibnamefont
  {Hennessy}}, \bibinfo {author} {\bibfnamefont {A.}~\bibnamefont {Badolato}},
  \bibinfo {author} {\bibfnamefont {M.}~\bibnamefont {Winger}}, \bibinfo
  {author} {\bibfnamefont {D.}~\bibnamefont {Gerace}}, \bibinfo {author}
  {\bibfnamefont {M.}~\bibnamefont {Atat{\"u}re}}, \bibinfo {author}
  {\bibfnamefont {S.}~\bibnamefont {Gulde}}, \bibinfo {author} {\bibfnamefont
  {S.}~\bibnamefont {F{\"a}lt}}, \bibinfo {author} {\bibfnamefont {E.~L.}\
  \bibnamefont {Hu}},\ and\ \bibinfo {author} {\bibfnamefont {A.}~\bibnamefont
  {Imamo{\u{g}}lu}},\ }\bibfield  {title} {\bibinfo {title} {Quantum nature of
  a strongly coupled single quantum dot--cavity system},\ }\href
  {https://doi.org/10.1038/nature05586} {\bibfield  {journal} {\bibinfo
  {journal} {Nature}\ }\textbf {\bibinfo {volume} {445}},\ \bibinfo {pages}
  {896} (\bibinfo {year} {2007})}\BibitemShut {NoStop}%
\bibitem [{\citenamefont {Pustovit}\ and\ \citenamefont
  {Shahbazyan}(2009)}]{PhysRevLett102.077401(2009)}%
  \BibitemOpen
  \bibfield  {author} {\bibinfo {author} {\bibfnamefont {V.~N.}\ \bibnamefont
  {Pustovit}}\ and\ \bibinfo {author} {\bibfnamefont {T.~V.}\ \bibnamefont
  {Shahbazyan}},\ }\bibfield  {title} {\bibinfo {title} {Cooperative emission
  of light by an ensemble of dipoles near a metal nanoparticle: The plasmonic
  dicke effect},\ }\href {https://doi.org/10.1103/PhysRevLett.102.077401}
  {\bibfield  {journal} {\bibinfo  {journal} {Phys. Rev. Lett.}\ }\textbf
  {\bibinfo {volume} {102}},\ \bibinfo {pages} {077401} (\bibinfo {year}
  {2009})}\BibitemShut {NoStop}%
\bibitem [{\citenamefont {Pustovit}\ and\ \citenamefont
  {Shahbazyan}(2010)}]{PhysRevB82.075429(2010)}%
  \BibitemOpen
  \bibfield  {author} {\bibinfo {author} {\bibfnamefont {V.~N.}\ \bibnamefont
  {Pustovit}}\ and\ \bibinfo {author} {\bibfnamefont {T.~V.}\ \bibnamefont
  {Shahbazyan}},\ }\bibfield  {title} {\bibinfo {title} {Plasmon-mediated
  superradiance near metal nanostructures},\ }\href
  {https://doi.org/10.1103/PhysRevB.82.075429} {\bibfield  {journal} {\bibinfo
  {journal} {Phys. Rev. B}\ }\textbf {\bibinfo {volume} {82}},\ \bibinfo
  {pages} {075429} (\bibinfo {year} {2010})}\BibitemShut {NoStop}%
\bibitem [{\citenamefont {Neuman}\ and\ \citenamefont
  {Aizpurua}(2018)}]{Optica10.1247(2018)}%
  \BibitemOpen
  \bibfield  {author} {\bibinfo {author} {\bibfnamefont {T.}~\bibnamefont
  {Neuman}}\ and\ \bibinfo {author} {\bibfnamefont {J.}~\bibnamefont
  {Aizpurua}},\ }\bibfield  {title} {\bibinfo {title} {Origin of the asymmetric
  light emission from molecular exciton-polaritons},\ }\href
  {https://doi.org/10.1364/OPTICA.5.001247} {\bibfield  {journal} {\bibinfo
  {journal} {Optica}\ }\textbf {\bibinfo {volume} {5}},\ \bibinfo {pages}
  {1247} (\bibinfo {year} {2018})}\BibitemShut {NoStop}%
\bibitem [{\citenamefont {Zengin}\ \emph {et~al.}(2015)\citenamefont {Zengin},
  \citenamefont {Wers\"all}, \citenamefont {Nilsson}, \citenamefont
  {Antosiewicz}, \citenamefont {K\"all},\ and\ \citenamefont
  {Shegai}}]{PhysRevLett114.157401(2015)}%
  \BibitemOpen
  \bibfield  {author} {\bibinfo {author} {\bibfnamefont {G.}~\bibnamefont
  {Zengin}}, \bibinfo {author} {\bibfnamefont {M.}~\bibnamefont {Wers\"all}},
  \bibinfo {author} {\bibfnamefont {S.}~\bibnamefont {Nilsson}}, \bibinfo
  {author} {\bibfnamefont {T.~J.}\ \bibnamefont {Antosiewicz}}, \bibinfo
  {author} {\bibfnamefont {M.}~\bibnamefont {K\"all}},\ and\ \bibinfo {author}
  {\bibfnamefont {T.}~\bibnamefont {Shegai}},\ }\bibfield  {title} {\bibinfo
  {title} {Realizing strong light-matter interactions between
  single-nanoparticle plasmons and molecular excitons at ambient conditions},\
  }\href {https://doi.org/10.1103/PhysRevLett.114.157401} {\bibfield  {journal}
  {\bibinfo  {journal} {Phys. Rev. Lett.}\ }\textbf {\bibinfo {volume} {114}},\
  \bibinfo {pages} {157401} (\bibinfo {year} {2015})}\BibitemShut {NoStop}%
\bibitem [{\citenamefont {Artuso}\ and\ \citenamefont
  {Bryant}(2008)}]{NanoLett8.2106(2008)}%
  \BibitemOpen
  \bibfield  {author} {\bibinfo {author} {\bibfnamefont {R.~D.}\ \bibnamefont
  {Artuso}}\ and\ \bibinfo {author} {\bibfnamefont {G.~W.}\ \bibnamefont
  {Bryant}},\ }\bibfield  {title} {\bibinfo {title} {Optical response of
  strongly coupled quantum dot-metal nanoparticle systems: Double peaked fano
  structure and bistability},\ }\href {https://doi.org/10.1021/nl800921z}
  {\bibfield  {journal} {\bibinfo  {journal} {Nano Lett.}\ }\textbf {\bibinfo
  {volume} {8}},\ \bibinfo {pages} {2106} (\bibinfo {year} {2008})}\BibitemShut
  {NoStop}%
\bibitem [{\citenamefont {Liu}\ \emph {et~al.}(2009)\citenamefont {Liu},
  \citenamefont {Langguth}, \citenamefont {Weiss}, \citenamefont {K{\"a}stel},
  \citenamefont {Fleischhauer}, \citenamefont {Pfau},\ and\ \citenamefont
  {Giessen}}]{NatMaterials8.758(2009)}%
  \BibitemOpen
  \bibfield  {author} {\bibinfo {author} {\bibfnamefont {N.}~\bibnamefont
  {Liu}}, \bibinfo {author} {\bibfnamefont {L.}~\bibnamefont {Langguth}},
  \bibinfo {author} {\bibfnamefont {T.}~\bibnamefont {Weiss}}, \bibinfo
  {author} {\bibfnamefont {J.}~\bibnamefont {K{\"a}stel}}, \bibinfo {author}
  {\bibfnamefont {M.}~\bibnamefont {Fleischhauer}}, \bibinfo {author}
  {\bibfnamefont {T.}~\bibnamefont {Pfau}},\ and\ \bibinfo {author}
  {\bibfnamefont {H.}~\bibnamefont {Giessen}},\ }\bibfield  {title} {\bibinfo
  {title} {Plasmonic analogue of electromagnetically induced transparency at
  the drude damping limit},\ }\href {https://doi.org/10.1038/nmat2495}
  {\bibfield  {journal} {\bibinfo  {journal} {Nat. Mater.}\ }\textbf {\bibinfo
  {volume} {8}},\ \bibinfo {pages} {758} (\bibinfo {year} {2009})}\BibitemShut
  {NoStop}%
\bibitem [{\citenamefont {Rempe}\ \emph {et~al.}(1987)\citenamefont {Rempe},
  \citenamefont {Walther},\ and\ \citenamefont
  {Klein}}]{PhysRevLett58.353(1987)}%
  \BibitemOpen
  \bibfield  {author} {\bibinfo {author} {\bibfnamefont {G.}~\bibnamefont
  {Rempe}}, \bibinfo {author} {\bibfnamefont {H.}~\bibnamefont {Walther}},\
  and\ \bibinfo {author} {\bibfnamefont {N.}~\bibnamefont {Klein}},\ }\bibfield
   {title} {\bibinfo {title} {Observation of quantum collapse and revival in a
  one-atom maser},\ }\href {https://doi.org/10.1103/PhysRevLett.58.353}
  {\bibfield  {journal} {\bibinfo  {journal} {Phys. Rev. Lett.}\ }\textbf
  {\bibinfo {volume} {58}},\ \bibinfo {pages} {353} (\bibinfo {year}
  {1987})}\BibitemShut {NoStop}%
\bibitem [{\citenamefont {Dovzhenko}\ \emph {et~al.}(2018)\citenamefont
  {Dovzhenko}, \citenamefont {Ryabchuk}, \citenamefont {Rakovich},\ and\
  \citenamefont {Nabiev}}]{Nanoscale10.3589(2018)}%
  \BibitemOpen
  \bibfield  {author} {\bibinfo {author} {\bibfnamefont {D.~S.}\ \bibnamefont
  {Dovzhenko}}, \bibinfo {author} {\bibfnamefont {S.~V.}\ \bibnamefont
  {Ryabchuk}}, \bibinfo {author} {\bibfnamefont {Y.~P.}\ \bibnamefont
  {Rakovich}},\ and\ \bibinfo {author} {\bibfnamefont {I.~R.}\ \bibnamefont
  {Nabiev}},\ }\bibfield  {title} {\bibinfo {title} {Light–matter interaction
  in the strong coupling regime: configurations, conditions, and
  applications},\ }\href {https://doi.org/10.1039/C7NR06917K} {\bibfield
  {journal} {\bibinfo  {journal} {Nanoscale}\ }\textbf {\bibinfo {volume}
  {10}},\ \bibinfo {pages} {3589} (\bibinfo {year} {2018})}\BibitemShut
  {NoStop}%
\bibitem [{\citenamefont {Würthner}\ \emph {et~al.}(2011)\citenamefont
  {Würthner}, \citenamefont {Kaiser},\ and\ \citenamefont
  {Saha-Möller}}]{AngChIntEd15.3376(2011)}%
  \BibitemOpen
  \bibfield  {author} {\bibinfo {author} {\bibfnamefont {F.}~\bibnamefont
  {Würthner}}, \bibinfo {author} {\bibfnamefont {T.~E.}\ \bibnamefont
  {Kaiser}},\ and\ \bibinfo {author} {\bibfnamefont {C.~R.}\ \bibnamefont
  {Saha-Möller}},\ }\bibfield  {title} {\bibinfo {title} {J-aggregates: From
  serendipitous discovery to supramolecular engineering of functional dye
  materials},\ }\href {https://doi.org/https://doi.org/10.1002/anie.201002307}
  {\bibfield  {journal} {\bibinfo  {journal} {Angew. Chem. Int. Ed.}\ }\textbf
  {\bibinfo {volume} {50}},\ \bibinfo {pages} {3376} (\bibinfo {year}
  {2011})}\BibitemShut {NoStop}%
\bibitem [{\citenamefont {Hammaker}\ \emph {et~al.}(1965)\citenamefont
  {Hammaker}, \citenamefont {Francis},\ and\ \citenamefont
  {Eischens}}]{SpectActa7.1295(1965)}%
  \BibitemOpen
  \bibfield  {author} {\bibinfo {author} {\bibfnamefont {R.}~\bibnamefont
  {Hammaker}}, \bibinfo {author} {\bibfnamefont {S.}~\bibnamefont {Francis}},\
  and\ \bibinfo {author} {\bibfnamefont {R.}~\bibnamefont {Eischens}},\
  }\bibfield  {title} {\bibinfo {title} {Infrared study of intermolecular
  interactions for carbon monoxide chemisorbed on platinum},\ }\href
  {https://doi.org/https://doi.org/10.1016/0371-1951(65)80213-2} {\bibfield
  {journal} {\bibinfo  {journal} {Spectrochim. Acta}\ }\textbf {\bibinfo
  {volume} {21}},\ \bibinfo {pages} {1295} (\bibinfo {year}
  {1965})}\BibitemShut {NoStop}%
\bibitem [{\citenamefont {Zirkelbach}\ \emph {et~al.}(2022)\citenamefont
  {Zirkelbach}, \citenamefont {Mirzaei}, \citenamefont {Deperasińska},
  \citenamefont {Kozankiewicz}, \citenamefont {Gurlek}, \citenamefont
  {Shkarin}, \citenamefont {Utikal}, \citenamefont {Götzinger},\ and\
  \citenamefont {Sandoghdar}}]{JChemPhys156.104301(2022)}%
  \BibitemOpen
  \bibfield  {author} {\bibinfo {author} {\bibfnamefont {J.}~\bibnamefont
  {Zirkelbach}}, \bibinfo {author} {\bibfnamefont {M.}~\bibnamefont {Mirzaei}},
  \bibinfo {author} {\bibfnamefont {I.}~\bibnamefont {Deperasińska}}, \bibinfo
  {author} {\bibfnamefont {B.}~\bibnamefont {Kozankiewicz}}, \bibinfo {author}
  {\bibfnamefont {B.}~\bibnamefont {Gurlek}}, \bibinfo {author} {\bibfnamefont
  {A.}~\bibnamefont {Shkarin}}, \bibinfo {author} {\bibfnamefont
  {T.}~\bibnamefont {Utikal}}, \bibinfo {author} {\bibfnamefont
  {S.}~\bibnamefont {Götzinger}},\ and\ \bibinfo {author} {\bibfnamefont
  {V.}~\bibnamefont {Sandoghdar}},\ }\bibfield  {title} {\bibinfo {title}
  {High-resolution vibronic spectroscopy of a single molecule embedded in a
  crystal},\ }\href {https://doi.org/10.1063/5.0081297} {\bibfield  {journal}
  {\bibinfo  {journal} {J. Chem. Phys.}\ }\textbf {\bibinfo {volume} {156}},\
  \bibinfo {pages} {104301} (\bibinfo {year} {2022})}\BibitemShut {NoStop}%
\bibitem [{\citenamefont {Chikkaraddy}\ \emph {et~al.}(2016)\citenamefont
  {Chikkaraddy}, \citenamefont {de~Nijs}, \citenamefont {Benz}, \citenamefont
  {Barrow}, \citenamefont {Scherman}, \citenamefont {Rosta}, \citenamefont
  {Demetriadou}, \citenamefont {Fox}, \citenamefont {Hess},\ and\ \citenamefont
  {Baumberg}}]{Nature535.127(2016)}%
  \BibitemOpen
  \bibfield  {author} {\bibinfo {author} {\bibfnamefont {R.}~\bibnamefont
  {Chikkaraddy}}, \bibinfo {author} {\bibfnamefont {B.}~\bibnamefont
  {de~Nijs}}, \bibinfo {author} {\bibfnamefont {F.}~\bibnamefont {Benz}},
  \bibinfo {author} {\bibfnamefont {S.~J.}\ \bibnamefont {Barrow}}, \bibinfo
  {author} {\bibfnamefont {O.~A.}\ \bibnamefont {Scherman}}, \bibinfo {author}
  {\bibfnamefont {E.}~\bibnamefont {Rosta}}, \bibinfo {author} {\bibfnamefont
  {A.}~\bibnamefont {Demetriadou}}, \bibinfo {author} {\bibfnamefont
  {P.}~\bibnamefont {Fox}}, \bibinfo {author} {\bibfnamefont {O.}~\bibnamefont
  {Hess}},\ and\ \bibinfo {author} {\bibfnamefont {J.~J.}\ \bibnamefont
  {Baumberg}},\ }\bibfield  {title} {\bibinfo {title} {Single-molecule strong
  coupling at room temperature in plasmonic nanocavities},\ }\href
  {https://doi.org/10.1038/nature17974} {\bibfield  {journal} {\bibinfo
  {journal} {Nature}\ }\textbf {\bibinfo {volume} {535}},\ \bibinfo {pages}
  {127} (\bibinfo {year} {2016})}\BibitemShut {NoStop}%
\bibitem [{\citenamefont {Santhosh}\ \emph {et~al.}(2016)\citenamefont
  {Santhosh}, \citenamefont {Bitton}, \citenamefont {Chuntonov},\ and\
  \citenamefont {Haran}}]{Santhosh2016}%
  \BibitemOpen
  \bibfield  {author} {\bibinfo {author} {\bibfnamefont {K.}~\bibnamefont
  {Santhosh}}, \bibinfo {author} {\bibfnamefont {O.}~\bibnamefont {Bitton}},
  \bibinfo {author} {\bibfnamefont {L.}~\bibnamefont {Chuntonov}},\ and\
  \bibinfo {author} {\bibfnamefont {G.}~\bibnamefont {Haran}},\ }\bibfield
  {title} {\bibinfo {title} {Vacuum rabi splitting in a plasmonic cavity at the
  single quantum emitter limit},\ }\href {https://doi.org/10.1038/ncomms11823}
  {\bibfield  {journal} {\bibinfo  {journal} {Nat. Comm.}\ }\textbf {\bibinfo
  {volume} {7}},\ \bibinfo {pages} {11823} (\bibinfo {year}
  {2016})}\BibitemShut {NoStop}%
\bibitem [{\citenamefont {Zhang}\ \emph {et~al.}(2017)\citenamefont {Zhang},
  \citenamefont {Meng}, \citenamefont {Zhang}, \citenamefont {Luo},
  \citenamefont {Yu}, \citenamefont {Yang}, \citenamefont {Zhang},
  \citenamefont {Esteban}, \citenamefont {Aizpurua}, \citenamefont {Luo},
  \citenamefont {Yang}, \citenamefont {Dong},\ and\ \citenamefont
  {Hou}}]{Zhang2017}%
  \BibitemOpen
  \bibfield  {author} {\bibinfo {author} {\bibfnamefont {Y.}~\bibnamefont
  {Zhang}}, \bibinfo {author} {\bibfnamefont {Q.-S.}\ \bibnamefont {Meng}},
  \bibinfo {author} {\bibfnamefont {L.}~\bibnamefont {Zhang}}, \bibinfo
  {author} {\bibfnamefont {Y.}~\bibnamefont {Luo}}, \bibinfo {author}
  {\bibfnamefont {Y.-J.}\ \bibnamefont {Yu}}, \bibinfo {author} {\bibfnamefont
  {B.}~\bibnamefont {Yang}}, \bibinfo {author} {\bibfnamefont {Y.}~\bibnamefont
  {Zhang}}, \bibinfo {author} {\bibfnamefont {R.}~\bibnamefont {Esteban}},
  \bibinfo {author} {\bibfnamefont {J.}~\bibnamefont {Aizpurua}}, \bibinfo
  {author} {\bibfnamefont {Y.}~\bibnamefont {Luo}}, \bibinfo {author}
  {\bibfnamefont {J.-L.}\ \bibnamefont {Yang}}, \bibinfo {author}
  {\bibfnamefont {Z.-C.}\ \bibnamefont {Dong}},\ and\ \bibinfo {author}
  {\bibfnamefont {J.~G.}\ \bibnamefont {Hou}},\ }\bibfield  {title} {\bibinfo
  {title} {Sub-nanometre control of the coherent interaction between a single
  molecule and a plasmonic nanocavity},\ }\href
  {https://doi.org/10.1038/ncomms15225} {\bibfield  {journal} {\bibinfo
  {journal} {Nat. Comm.}\ }\textbf {\bibinfo {volume} {8}},\ \bibinfo {pages}
  {15225} (\bibinfo {year} {2017})}\BibitemShut {NoStop}%
\bibitem [{\citenamefont {Le~Gall}\ \emph {et~al.}(1997)\citenamefont
  {Le~Gall}, \citenamefont {Olivier},\ and\ \citenamefont
  {Greffet}}]{PhysRevB.55.10105(1997)}%
  \BibitemOpen
  \bibfield  {author} {\bibinfo {author} {\bibfnamefont {J.}~\bibnamefont
  {Le~Gall}}, \bibinfo {author} {\bibfnamefont {M.}~\bibnamefont {Olivier}},\
  and\ \bibinfo {author} {\bibfnamefont {J.-J.}\ \bibnamefont {Greffet}},\
  }\bibfield  {title} {\bibinfo {title} {Experimental and theoretical study of
  reflection and coherent thermal emissionby a sic grating supporting a
  surface-phonon polariton},\ }\href
  {https://doi.org/10.1103/PhysRevB.55.10105} {\bibfield  {journal} {\bibinfo
  {journal} {Phys. Rev. B}\ }\textbf {\bibinfo {volume} {55}},\ \bibinfo
  {pages} {10105} (\bibinfo {year} {1997})}\BibitemShut {NoStop}%
\end{thebibliography}%

\end{document}